\documentclass[a4paper,fleqn,usenatbib]{mnras}

\usepackage{newtxtext,newtxmath}
\usepackage[T1]{fontenc}
\usepackage{ae,aecompl}
\usepackage{graphicx}	% Including figure files
\usepackage{enumitem}
\usepackage{rotating}

\newcommand{\xmm}{{\sc{XMM}}\emph{-Newton}}
\newcommand{\te}{\emph{TESS}}

\title[X-ray detections from WRs]{New X-ray detections of known Wolf-Rayet stars\thanks{Based on data obtained with \xmm , an ESA Science Mission with instruments and contributions directly funded by ESA Member States and the USA (NASA).}}

\author[Y. Naz\'e et al.]{Ya\"el~Naz\'e\thanks{F.R.S.-FNRS Senior Research Associate, email: ynaze@uliege.be}, Eric Gosset\thanks{F.R.S.-FNRS Research Director}, and Quentin Marechal
\\
Groupe d'Astrophysique des Hautes Energies, STAR, Universit\'e de Li\`ege, Quartier Agora (B5c, Institut d'Astrophysique et de G\'eophysique), \\
All\'ee du 6 Ao\^ut 19c, B-4000 Sart Tilman, Li\`ege, Belgium
}

\pubyear{2020}

\begin{document}
\label{firstpage}
\pagerange{\pageref{firstpage}--\pageref{lastpage}}
\maketitle

\begin{abstract}
Using \xmm, we undertook a dedicated project to search for X-ray bright wind-wind collisions in 18 WR+OB systems. We complemented these observations with {\it Swift} and {\it Chandra} datasets, allowing for the study of two additional systems. We also improved the ephemerides, for these systems displaying photometric changes, using \te, {\it Kepler}, and {\it ASAS-SN} data. Five systems displayed a very faint X-ray emission ($\log [L_{\rm X}/L_{\rm BOL}]<-8$) and three a faint one ($\log [L_{\rm X}/L_{\rm BOL}]\sim-7$), incompatible with typical colliding wind emission: not all WR binaries are thus X-ray bright. In a few other systems, X-rays from the O-star companion cannot be excluded as being the true source of X-rays (or a large contributor). In two additional cases, the emission appears faint but the observations were taken with the WR wind obscuring the line-of-sight, which could hide a colliding wind emission. Clear evidence of colliding winds was however found in the remaining six systems (WR\,19, 21, 31, 97, 105, 127). In WR\,19, increased absorption and larger emission at periastron are even detected, in line with expectations of adiabatic collisions.
\end{abstract}

\begin{keywords}
stars: early-type -- X-ray: stars -- stars: massive -- stars: Wolf-Rayet
\end{keywords}

\section{Introduction}
Massive stars blow strong winds, making them important feedback agents. However, the exact wind properties (mass-loss rates, for example) remain debated. One way to constrain the wind characteristics is to study colliding wind binaries, i.e. systems composed of two massive stars whose dense winds collide, generally at high speeds. Such collisions may result in prominent X-ray emissions (for a review, see \citealt{rau16}). This is especially the case for WR+OB binaries as single WRs are X-ray faint ($\log [L_{\rm X}/L_{\rm BOL}]<-7$), with many non-detections reported in literature \citep{gos05,osk03,ski06}.

The diagnostic value of X-rays associated to wind-wind collisions lies in their variations: the line-of-sight absorption changes as the different winds (the dense one of the WR, or the more tenuous one of its companion) interpose between the collision zone and the observer; the intrinsic strength of the collision varies with the orbital separation in eccentric systems. When the orbital parameters are known, the observed modulations directly constrain the wind parameters (mass-loss rate, wind speed) as has been shown in previous studies (e.g. \citealt{lom15,gos16,aro19}). The overall luminosity also depends on wind parameters (see \citealt{zhe12}). 

Despite two decades of massive star studies at high energies, few WR+OB binaries have been studied in detail. Often, analyses have focused on the systems detected in previous X-ray surveys such as the {\it ROSAT All-Sky Survey}. In the meantime, the sensitivity of X-ray facilities has enormously improved, but still no new survey is available yet. To provide a larger census allowing to probe different physical conditions, we perform a dedicated detection experiment with \xmm, covering a large interval of wind, stellar, and orbital parameters. This paper reports on the results of this project. Section 2 presents the (X-ray and optical) data along with their reduction, Section 3 explains the derivation of the hot plasma properties, which are then discussed in Section 4. Section 5 summarizes and concludes the results.

\begin{table*}
  \centering
  \scriptsize
\caption{Journal of the X-ray observations and properties of the targets.}
\label{journal}
\setlength{\tabcolsep}{3.3pt}
\begin{tabular}{llrlccccccccc}
  \hline\hline
  ObsID & mid-HJD & $T_{\rm exp}$ (Fl?) & Target & sp. type & d(pc) & $E(b-v)$ & $\log(L^{\rm WR}_{\rm BOL}$ & $\log(\dot M)$ & $v_{\infty}$ & $P$(d) & $T_0$ & $\phi$ \\
  &--2\,450\,000. & (ks) & & & & & $/L_{\odot})$\\
  \hline
0840210201 &8908.10366 &7.6,8.0,4.3-y     &WR\,4    &WC5$^+$        & 3711 &0.6$^+$  &5.71$^+$ &--4.37$^+$ &2528$^+$ &2.4096 [1] & 2446620.444 [1] & 0.46 \\
0840210501 &8766.44899 &11.2,11.8,5.5-y   &WR\,9    &WC5+O7$^*$     & 4659 &1.12$^*$ &5.6$^-$  &--4.39$^-$ &2780$^-$ &14.305 [2] & 2446036.075 [2] & 0.92 \\
0840210601 &8830.27711 &20.1,20.3,16.6-n  &WR\,12   &WN8h + OB$^+$  & 5916 &0.8$^+$  &5.98$^+$ &--4.30$^+$ &1200$^+$ &23.92336[3]& 2458903.544 [0] & 0.94 \\
0780070401 &7504.14059 &4.4,4.4,1.3-n     &WR\,14   &WC7$^+$        & 2228 &0.65$^+$ &5.78$^+$ &--4.39$^+$ &2194$^+$ &2.42$^*$   &                 &     \\
0780070801 &7526.90709 &6.5,6.4,4.3-n     &         &               &      &         &         &           &         &           &                 & +0.41 after \\
0792382601 &7612.36184 &5.6,5.3,1.3-y     &WR\,19   &WC4pd+O9.6$^*$ & 4494 &1.33$^*$ &5.2$^-$  &--4.67$^-$ &3310$^-$ &3689. [4]  & 2450500.5 [4]   & 0.93 \\
0793183301 &7725.03849 &8.3,8.4,5.1-y     &         &               &      &         &         &           &         &           &                 & 0.96 \\
0793183701 &7793.83264 &10.7,10.2,7.0-y   &         &               &      &         &         &           &         &           &                 & 0.98 \\
0804050201 &7901.81438 &63.6,63.8,55.0-n  &         &               &      &         &         &           &         &            &                 &0.01 \\
00034544002&7533.11794 &1.4               &         &               &      &         &         &           &         &            &                 &0.91 \\
00034544005&7602.94421 &4.0               &         &               &      &         &         &           &         &            &                 &0.93 \\
0780070701 &7616.35955 &6.6,6.5,4.4-n     &WR\,21   &WN5 + O4-6$^+$ & 4122 &0.56$^*$ &5.53$^-$ &--4.68$^-$ &1485$^-$ &8.25443 [3] & 2458591.912 [0] &0.81 \\
0800000301 &7915.43058 &11.5,11.5,8.7-n   &WR\,29   &WN7h+O$^*$     & 6318 &0.85$^*$ &5.73$^-$ &--4.52$^-$ &1369$^-$ &3.16412 [5] & 2458621.212 [0] &0.94 \\
0840210801 &8852.80223 &14.7,14.7,11.2-n  &WR\,30   &WC6+O6-8$^*$   & 5326 &0.62$^*$ &5.7$^-$  &--4.47$^-$ &2100$^*$ &18.827 [0]  & 2458316.978 [0] &0.46 \\
0840210401 &8844.25912 &8.8,8.8,5.2-y     &WR\,31   &WN4 + O8V$^+$  &10309 &0.55$^*$ &5.66$^-$ &--4.66$^-$ &1713$^-$ &4.830657 [3]& 2458619.841 [0] &0.46 \\
0780070101 &7736.52140 &8.2,8.2,5.9-n     &WR\,42   &WC7+O7V$^*$    & 2513 &0.3$^*$  &5.6$^-$  &--4.57$^-$ &1500$^*$ &7.752 [0]   & 2458616.859 [0] &0.44 \\
0780070501 &7614.33673 &9.6,9.3,5.8-y     &WR\,69   &WC9d$^+$       & 3509 &0.55$^+$ &5.33$^+$ &--4.87$^+$ &1089$^+$ &2.293$^*$   &                 &     \\
0840210701 &8910.10163 &8.5,8.7,3.6-n     &WR\,97   &WN5+O7$^*$     & 2001 &0.97$^*$ &5.53$^-$ &--4.68$^-$ &1900$^*$ &12.595 [6]  & 2444903.354 [6] &0.09 \\
00010064002&7865.46058 &3.7               &         &               &      &         &        &           &         &            &                 &0.15 \\
0840211101 &8941.89666 &19.1,18.6,12.0-y  &WR\,98   &WN7/C+O8-9 [7] & 2028 &1.6$^+$  &5.63$^+$ &--4.57$^-$ &1600$^+$ &47.825 [7]  & 2445676.4 [7]   &0.38 \\
9976       &5048.68022 &19.7              &WR\,98a  &WC8-9vd+? [8]  & 1361 &3.36$^*$ &5.5$^+$  &--4.60$^+$ &1600$^+$ & 564 [8]    &                 &     \\
0780070201 &7815.17265 &12.1,11.8,7.8-y   &WR\,103  &WC9d$^+$       & 3619 &0.52$^+$ &5.50$^+$ &--4.56$^+$ &1190$^+$ &1.75404 [9] & 2442862.87 [9]  &0.49 \\
00043802001&6207.48861 &0.5               &WR\,105  &WN9h$^+$       & 1760 &2.15$^+$ &5.89$^+$ &--4.40$^+$ &800$^+$  &            &                 &     \\
00049689001&6449.91829 &0.8 \\
00049689002&8537.79152 &1.1 \\
00049689003&8582.47982 &1.1 \\
0780070601 &7674.18646 &9.9,9.9,7.4-n     &WR\,113  &WC8d+O8-9IV$^*$& 1819 &0.79$^*$ &5.6$^-$  &--4.53$^-$ &1700$^*$ &29.7 [10]   & 2443429.5  [10] &0.62 \\
0840210101 &8760.24680 &11.4,11.4,7.1-$^a$&WR\,127  &WN3 + O9.5V$^+$& 3120 &0.41$^*$ &5.55$^-$ &--5.33$^-$ &2333$^-$ &9.5465 [0]  & 2458427.015 [0] &0.91 \\
0840210301 &8760.47373 &16.5,16.4,9.9-y   &WR\,128  &WN4(h)-w$^+$   & 2925 &0.32$^+$ &5.22$^+$ &--5.40$^+$ &2050$^+$ &3.85 [11]   & 2444788.65 [11] &0.05 \\
0780070301 &7553.96962 &15.8,15.8,12.7-n  &WR\,153ab&WN6/WCE+O6I$^*$& 4146 &0.56$^*$ &5.43$^-$ &--4.57$^-$ &1785$^*$ &6.6887[12]  & 2458786.387 [0] &0.75 \\
           &           &                  &         &+ B0I+B1V-III  &      &         &        &           &         &3.4663[12]  & 2458787.080 [0] &0.26 \\
  \hline      
\end{tabular}

{\scriptsize Effective exposure times are provided for each instrument for XMM-EPIC, in the order MOS1, MOS2, pn, along with a yes/no flag indicating the presence of flares; $^a$ for WR\,127, no clear flare is seen for MOS but one was cut for pn; finally, some exposures display no isolated flares but elevated background during their entire duration (0780070401, 0840210201/601/701). For stellar properties, $^*$ indicates a value taken from \citet{vdh01},  $^+$ indicates values taken from \citet{ham19} for WN stars or \citet{san19} for WC stars (for WR\,98, values of the pseudo-fit of \citet{san12} are used, taking into account the difference in distance for the luminosity),  $^-$ indicates mean values for stars of the same spectral types analyzed by \citet{ham19} for WN stars or \citet{san19} for WC and WN/WCE stars.  Distances come from \citet{bai18}, $T_0$ correspond to a conjunction with the WR star in front, except for WR\,19 for which it is a periastron passage.  
Number within squared brackets indicate references for the ephemerides: [0] this work, [1] \citet{rus89}, [2] \citet{bar01}, [3] \citet{fah12}, [4] \citet{wil09}, [5] \citet{gam09}, [6] \citet{lam96}, [7] \citet{gam02}, [8] \citet{wil03}, [9] \citet{mof86}, [10] \citet{hil18}, [11] \citet{ant82}, [12] \citet{dem02}. } 
\end{table*}

\section{Data reduction and analysis}
\subsection{\xmm\ }
To select our targets, we have used the 7th catalog of WR stars of \citet{vdh01}. We have considered only his secure cases of WR binaries (i.e. spectroscopic binaries - SB1 or SB2 - with a known orbital period in Tables 18 and 19 of the catalog). There are 41 objects with these criteria: 17 have already been observed by \xmm\ or {\it Chandra} at least once and three other ones are too faint or absorbed ($V>14$, $A_v>7$) for the proposed detection experiment. This leaves 21 suitable targets, of which 17 were observed by \xmm\ in the framework of our dedicated WR survey (programs 078007, 080000, and 084021 - PI Naz\'e). An additional system (WR\,19) has archived observations (0792382601, 0793183301/701, 0804050201 - PIs N. Schartel or Y. Sugawara). The list of the exposures is available in Table \ref{journal}. This table also provides information on the stellar properties, with distances from \citet{bai18}, stellar properties from \citet{vdh01}, \citet{ham19}, or \citet{san19}, and orbital ephemerides from literature or this work (see Appendix). 

The \xmm\ data were processed with the Science Analysis Software (SAS) v18.0.0 using calibration files available in Jan. 2020 and following the recommendations of the \xmm\ team\footnote{SAS threads, see \\ http://xmm.esac.esa.int/sas/current/documentation/threads/ }. After the initial pipeline processing, the European Photon Imaging Camera (EPIC) observations were filtered to keep only the best-quality data ({\sc{pattern}} 0--12 for MOS and 0--4 for pn). Light curves for events beyond 10\,keV were built for the full fields. Using them, background flares were detected in about half of the exposures and the time intervals corresponding to isolated flaring events were discarded before further processing.

To assess the detection of our targets and the crowding near them (in order to choose the best extraction region), a source detection was performed on each EPIC dataset using the task {\it edetect\_chain}, which uses first sliding box algorithms and then performs a PSF fitting, on the 0.3--10.0\,keV energy band and for a log-likelihood of 10. We obtained a formal detection for twelve targets and their EPIC count rates are provided in Table \ref{det}. For the non-detections, we derived upper values on count rates from the sensitivity map corresponding to a log-likelihood of 2.3 (i.e. a threshold such that sources with these count rate levels would be detected 90\% of the time at these sky positions)\footnote{While a 90\% limit is classically used, one may prefer to derive the sensitivity map using the same likelihood as for detection: in this case, the derived count rates would be $\sim$3 times (or $\sim$0.5\,dex) larger.}; they are listed in Table \ref{det}.

\begin{table}
\centering
  \scriptsize
\caption{X-ray count rates (with their 1$\sigma$ errors) or 90\% upper limits for our WR sample in the 0.3--10.0\,keV energy band.}
\label{det}
\setlength{\tabcolsep}{3.3pt}
\begin{tabular}{llccc}
  \hline\hline
  ObsID & Name & \multicolumn{3}{c}{Count rates (cts\,s$^{-1}$)}\\
  &      &     pn or XRT & MOS1 & MOS2 \\
  \hline
\multicolumn{5}{l}{\it Non-detections}\\
0840210201 & WR\,4   & $<$3.2e-3 & $<$1.0e-3 & $<$6.9e-4 \\
0840210601 & WR\,12  & $<$1.0e-3 & $<$5.2e-4 & $<$3.4e-4 \\
0780070401 & WR\,14  & $<$9.5e-3 & $<$2.0e-3 & $<$1.5e-3 \\ 
0780070801 &         & $<$1.3e-3 & $<$8.9e-4 & $<$9.8e-4 \\
0780070101 & WR\,42  & $<$1.0e-3 & $<$3.7e-4 & $<$3.7e-4 \\
0780070501 & WR\,69  & $<$1.0e-3 & $<$3.1e-4 & $<$3.3e-4 \\ 
0780070201 & WR\,103 & $<$7.1e-4 & $<$5.0e-4 & $<$5.2e-4 \\
  \hline
\multicolumn{5}{l}{\it Detections}\\
0840210501 & WR\,9     & (1.11$\pm$0.22)e-2 & (3.15$\pm$0.88)e-3  & (2.58$\pm$0.68)e-3 \\
0792382601 & WR\,19    & (1.00$\pm$0.42)e-2 & (3.32$\pm$1.12)e-3  & (4.45$\pm$1.27)e-3 \\
0793183301 &           & (9.40$\pm$1.90)e-3 & (3.83$\pm$1.03)e-3  & (2.93$\pm$0.82)e-3 \\
0793183701 &           & (1.31$\pm$0.19)e-2 & (4.55$\pm$0.94)e-3  & (4.82$\pm$0.90)e-3 \\
0804050201 &           & (1.81$\pm$0.07)e-2 & (6.98$\pm$0.45)e-3  & (7.04$\pm$0.42)e-3 \\
00034544002&           & (5.2$\pm$2.6)e-3\\
00034544005&           & (1.2$\pm$0.8)e-3\\
0780070701 & WR\,21    & (6.17$\pm$0.46)e-2 & (2.54$\pm$0.28)e-2  & (2.41$\pm$0.24)e-2 \\
0800000301 & WR\,29    & (7.00$\pm$1.25)e-3 & (3.04$\pm$0.80)e-3  & (2.40$\pm$0.64)e-3 \\
0840210801 & WR\,30    & (3.32$\pm$0.85)e-3 & (1.94$\pm$0.61)e-3  & (7.16$\pm$3.70)e-4 \\
0840210401 & WR\,31    & (6.74$\pm$0.46)e-2 & (2.45$\pm$0.23)e-2  & (2.52$\pm$0.21)e-2 \\
0840210701 & WR\,97    & (4.43$\pm$0.16)e-1 & (1.50$\pm$0.06)e-1  & (1.70$\pm$0.06)e-1 \\
00010064002&           & (3.9$\pm$0.4)e-2\\
0840211101 & WR\,98    & (3.75$\pm$0.90)e-3 & (1.55$\pm$0.47)e-3  & (3.18$\pm$0.57)e-3 \\
9976       & WR\,98a   & (5.64$\pm$0.61)e-3\\
00043802001& WR\,105   & (1.6$\pm$0.9)e-2\\
00049689001&           & (1.2$\pm$0.5)e-2\\
00049689002&           & (7.0$\pm$3.5)e-3\\
00049689003&           & (1.6$\pm$0.4)e-2\\
0780070601 & WR\,113   & (1.15$\pm$0.05)e-1 & (3.64$\pm$0.25)e-2  & (4.14$\pm$0.25)e-2 \\
0840210101 & WR\,127   & (1.36$\pm$0.05)e-1 & (4.65$\pm$0.26)e-2  & (4.71$\pm$0.25)e-2 \\
0840210301 & WR\,128   & (2.09$\pm$0.20)e-2 & (5.63$\pm$0.85)e-3  & (5.29$\pm$0.76)e-3 \\
0780070301 & WR\,153ab & (6.05$\pm$0.31)e-2 & (2.12$\pm$0.15)e-2  & (2.02$\pm$0.14)e-2 \\
  \hline      
\end{tabular}
\end{table}

For the brighter detections (i.e. EPIC-pn count rate larger than $\sim$0.01\,cts\,s$^{-1}$), we then extracted EPIC spectra using circular regions centered on the Simbad positions of the targets with radii of 30\arcsec\ (25\arcsec\ for WR\,128). Background was derived in nearby circular regions of 30--50\arcsec\ radius devoid of sources. Spectra and their dedicated calibration matrices (ancillary response file and redistribution matrix file, which are used to calibrate the flux and energy axes, respectively) were derived using the task {\it{especget}}. EPIC spectra were grouped with {\it{specgroup}} to obtain an oversampling factor of five and to ensure that a minimum signal-to-noise ratio of 3 (i.e., a minimum of ten counts) was reached in each spectral bin of the background-corrected spectra. 

\subsection{{\it Swift} }

We searched the {\it Swift} archives for observations of our targets. We found exposures of WR\,19 and 97 which complement the \xmm\ data, as well as a detection of WR\,105 (Table \ref{journal}). The latter star is a known non-thermal radio emitter \citep{mon09}: such a feature is known to arise in wind-wind collisions \citep{van06}, which explains its inclusion here. Count rates in the 0.3--10\,keV band, extracted using the online tool\footnote{https://www.swift.ac.uk/user\_objects/}, are reported in Table \ref{det}. For WR\,97, a single observation is available and the associated spectrum was extracted using the online tool. For WR\,19 and 105, several observations are available but the count rates did not appear to significantly vary ($\chi^2$ test with 1\% significance level) hence exposures could in principle be combined. However, the signal-to-noise ratio remained too poor, even after combining, to get usable spectra. Nevertheless, because of the absence of significant variation, the weighted mean of WR\,105 count rate was calculated and used to estimate its X-ray flux (see end of Sect. 3 and Table \ref{conv}).

\subsection{\it Chandra}
We also examined the {\it Chandra} archives for observations of our targets and we found a detection of WR\,98a. This object is a non-thermal radio emitter \citep{mon02} and an episodic dust maker surrounded by a pinwheel nebula \citep[and references therein]{wil03}: the presence of colliding winds in this system is therefore known. Within CIAO v4.9, we extracted the source and background spectra using {\sc specextract}. As WR\,98a appeared far off-axis, a circle of 20\,px was used for the source region, the background was defined by the surrounding annulus with external radius of 40\,px, and weighted response matrices were calculated. A binning similar to that done for \xmm\ and {\it Swift} data was used.

\section{X-ray properties}
For each target, all available EPIC spectra were fitted simultaneously within Xspec v12.9.1p. The fitting was similar for EPIC, ACIS, XRT spectra: as is usual for massive stars, we used absorbed optically thin thermal plasma models. The first absorption component was fixed to the interstellar column, derived from the known color excess (Table \ref{journal}) using the formula of \citet[$N_{\rm H}^{\rm ISM}=6.12\times 10^{21}\times E(B-V)$\,cm$^{-2}$]{gud12} and the conversion factor $E(B-V)=1.21\times E(b-v)$ \citep{vdh01}. The second absorbing component accounts for possible local absorption and was allowed to vary. In one case (WR\,128), however, the additional absorption converged to $\sim$0 and yielded erratic results with unrealistic errors, hence we fixed it to zero. 

\begin{table*}
%\begin{sidewaystable*}
\centering
  \scriptsize
\caption{Results of the spectral fitting (see text for details). }
\label{fits}
\scriptsize
\setlength{\tabcolsep}{1pt}
\begin{tabular}{lccccccccccccc}
\hline\hline
Name                    & \# Cts & $N_{\rm H}^{\rm ISM}$ &$N_{\rm H}$ & $kT_1$ & $norm_1$ &$ kT_2$ & $norm_2$ & $\chi^2_{\nu}$(dof) & $F_{\rm X}^{\rm obs}$(tot$^a$) & $L_{\rm X}^{\rm ISM\,cor}$(tot) & $\log(L_{\rm X}^{\rm ISM\,cor}({\rm tot})$ & $HR$ & $L_{\rm X}^{\rm full\,abs\,cor}$(tot) \\
                        & & \multicolumn{2}{c}{($10^{22}$\,cm$^{-2}$)} & (keV) & (cm$^{-5}$) & (keV) & (cm$^{-5}$) & & (erg\,cm$^{-2}$\,s$^{-1}$) & (erg\,s$^{-1}$) & $/L^{\rm WR}_{\rm BOL})$ & & (erg\,s$^{-1}$) \\
\hline
\multicolumn{14}{l}{\it solar abundances $phabs\times phabs\times \sum apec$}\\
WR\,19$^d$&2745&0.98 &4.31$\pm$0.68 & 2.84$\pm$0.42 & (2.52$\pm$0.52)e-4&              &                    &1.21(96) & (1.14$\pm$0.08)e-13 & (3.00$\pm$0.21)e+32 & --6.31$\pm$0.03 &15.74$\pm$2.48 & (8.15$\pm$0.57)e32 \\ 
WR\,21   &594 &0.41 &0.33$\pm$0.27 & 4.46$\pm$1.60 & (2.53$\pm$0.47)e-4&              &                    &1.18(41) & (2.86$\pm$0.28)e-13 & (6.65$\pm$0.65)e+32 & --6.29$\pm$0.04 & 2.43$\pm$0.31 & (8.05$\pm$0.79)e32 \\ 
WR\,31   &917 &0.41 &0.44$\pm$0.23 & 6.64$\pm$3.43 & (2.60$\pm$0.36)e-4&              &                    &0.90(51) & (3.40$\pm$0.43)e-13 & (4.76$\pm$0.60)e+33 & --5.57$\pm$0.05 & 3.64$\pm$0.52 & (5.77$\pm$0.73)e33 \\ 
WR\,97   &8458&0.72 &1.04$\pm$0.09 & 0.76$\pm$0.08 & (1.59$\pm$0.33)e-3& 2.16$\pm$0.18& (1.87$\pm$0.16)e-3 &1.23(148)& (1.35$\pm$0.07)e-12 & (9.01$\pm$0.47)e+32 & --6.16$\pm$0.02 & 1.05$\pm$0.06 & (2.80$\pm$0.15)e33 \\ 
WR\,97$^s$&132 &0.72&0.66$\pm$0.71 & 1.99$\pm$1.31 & (2.29$\pm$1.13)e-3&              &                    &0.65(9)  & (1.28$\pm$0.33)e-12 & (8.15$\pm$2.10)e+32 & --6.20$\pm$0.11 & 1.34$\pm$0.33 & (1.36$\pm$0.35)e33 \\
WR\,98a  &135 &2.49 &0.64$\pm$0.88 & 1.83$\pm$0.58 & (2.63$\pm$1.28)e-4&              &                    &0.66(9)  & (9.18$\pm$2.58)e-14 & (4.23$\pm$1.19)e+31 & --7.46$\pm$0.12 & 1.15$\pm$0.34 & (7.25$\pm$2.04)e31 \\
WR\,113  &1675&0.59 &3.61$\pm$0.38 & 1.64$\pm$0.17 & (1.82$\pm$0.26)e-3&              &                    &1.04(99) & (4.67$\pm$0.25)e-13 & (2.04$\pm$0.11)e+32 & --6.87$\pm$0.02 & 5.37$\pm$0.57 & (9.19$\pm$0.49)e32 \\ 
WR\,127  &1931&0.30 &3.11$\pm$0.35 &0.088$\pm$0.004& (10.9$\pm$12.2)   & 2.75$\pm$0.29& (1.19$\pm$0.15)e-3 &1.10(151)& (6.41$\pm$0.43)e-13 & (7.84$\pm$0.53)e+32 & --6.24$\pm$0.03 & 6.41$\pm$0.96 & (7.20$\pm$0.48)e35 \\ 
WR\,128  &674 &0.24 &0 (fixed)     & 2.14$\pm$0.22 & (3.80$\pm$0.44)e-5&              &                    &2.82(19) & (3.74$\pm$0.49)e-14 & (4.83$\pm$0.63)e+31 & --7.12$\pm$0.06 & 0.66$\pm$0.09 & (4.83$\pm$0.63)e31 \\ 
WR\,153ab&2530&0.41 &0.87$\pm$0.11 & 0.40$\pm$0.10 & (6.13$\pm$3.04)e-4& 5.92$\pm$2.81& (7.36$\pm$2.74)e-5 &1.24(54) & (1.53$\pm$1.00)e-13 & (4.53$\pm$2.96)e+32 & --6.36$\pm$0.28 & 0.59$\pm$0.29 & (2.28$\pm$1.49)e33 \\ 
\multicolumn{14}{l}{\it WC abundances with He/H=100 solar, $phabs\times vphabs\times \sum vapec$}\\
WR\,19$^d$&&0.98 & (6.11$\pm$0.98)e-3& 3.68$\pm$0.73 & (1.66$\pm$0.21)e-6 &               &                    &0.97(96) & (1.21$\pm$0.09)e-13 & (3.17$\pm$0.24)e+32 & --6.28$\pm$0.03 &17.17$\pm$2.33 & (6.48$\pm$0.48)e32 \\
WR\,98a  &&2.49 & (1.11$\pm$1.23)e-3& 1.51$\pm$0.51 & (2.30$\pm$1.02)e-6 &               &                    &0.57(9)  & (8.62$\pm$2.54)e-14 & (3.97$\pm$1.17)e+31 & --7.48$\pm$0.13 & 1.14$\pm$0.30 & (7.83$\pm$2.31)e31 \\
WR\,113  &&0.59 & (4.49$\pm$0.62)e-3& 1.79$\pm$0.20 & (1.03$\pm$0.16)e-5 &               &                    &1.12(99) & (4.62$\pm$0.28)e-13 & (2.01$\pm$0.12)e+32 & --6.88$\pm$0.03 & 5.20$\pm$0.46 & (6.02$\pm$0.36)e32 \\
WR\,153ab&&0.41 & (5.26$\pm$0.96)e-4& 0.58$\pm$0.13 & (1.63$\pm$0.51)e-6 & 28.9$\pm$24.1 & (5.24$\pm$2.17)e-7 &1.60(54) & (1.49$\pm$0.82)e-13 & (4.53$\pm$2.49)e+32 & --6.36$\pm$0.24 & 0.55$\pm$0.21 & (9.34$\pm$5.14)e32 \\
\multicolumn{14}{l}{\it WNE abundances with He/H=100 solar, $phabs\times vphabs\times \sum vapec$}\\                                                                                             
WR\,21   &&0.41 & (5.63$\pm$5.86)e-3& 4.83$\pm$1.56 & (8.51$\pm$1.35)e-6 &               &                    &1.20(41) & (2.89$\pm$0.29)e-13 & (6.71$\pm$0.67)e+32 & --6.29$\pm$0.04 & 2.43$\pm$0.28 & (7.79$\pm$0.78)e32 \\
WR\,31   &&0.41 & (9.26$\pm$5.00)e-3& 6.90$\pm$3.97 & (9.11$\pm$1.02)e-6 &               &                    &0.91(51) & (3.40$\pm$0.69)e-13 & (4.74$\pm$0.96)e+33 & --5.57$\pm$0.09 & 3.68$\pm$0.55 & (5.66$\pm$1.15)e33 \\
WR\,97   &&0.72 & (2.21$\pm$0.21)e-2& 0.77$\pm$0.08 & (4.96$\pm$1.02)e-5 & 2.24$\pm$0.25 & (6.03$\pm$0.57)e-5 &1.25(148)& (1.34$\pm$0.07)e-12 & (8.91$\pm$0.47)e+32 & --6.16$\pm$0.02 & 1.06$\pm$0.06 & (2.53$\pm$0.13)e33 \\
WR\,97$^s$&&0.72& (9.66$\pm$19.8)e-3& 2.29$\pm$1.11 & (6.86$\pm$3.89)e-5 &               &                    &0.67(9)  & (1.33$\pm$0.31)e-12 & (8.43$\pm$1.97)e+32 & --6.19$\pm$0.10 & 1.38$\pm$0.32 & (1.20$\pm$0.28)e33 \\
WR\,127  &&0.30 & (7.07$\pm$0.87)e-2&0.087$\pm$0.002& (0.32$\pm$0.30)    & 2.98$\pm$0.34 & (3.69$\pm$0.41)e-5 &1.14(151)& (6.45$\pm$0.33)e-13 & (7.86$\pm$0.40)e+32 & --6.24$\pm$0.02 & 6.55$\pm$0.88 & (4.71$\pm$2.41)e35 \\
WR\,128  &&0.24 &0 (fixed)          & 2.10$\pm$0.23 & (1.31$\pm$0.15)e-6 &               &                    &2.84(19) & (3.71$\pm$0.58)e-14 & (4.80$\pm$0.75)e+31 & --7.12$\pm$0.07 & 0.65$\pm$0.09 & (4.80$\pm$0.75)e31 \\
\hline
\end{tabular}

{\scriptsize The second column in the top part provides the number of spectral data counts in the source region (sum of all EPIC spectra for \xmm\ data).  $^a$ The total band (tot) corresponds to the 0.5--10.0\,keV energy band. $HR$ is defined as the ratio between the ISM-corrected fluxes in the hard (2.0--10.0\,keV) and soft (0.5--2.0\,keV) energy bands. $^d$ The WR\,19 spectra refer to observation with ObsID=0804050201. $^s$ This suffix indicates the {\it Swift} spectrum. Note that all errors are 1$\sigma$ errors, with those on luminosities only reflecting the flux errors (uncertainties on distance or model are not considered).}

%\end{sidewaystable*}
\end{table*}

Abundances cannot be determined from short, low-resolution snapshots such as the ones used in this paper hence they were fixed. Fitting was first performed using solar abundances of \citet{asp09} for all components ($phabs\times phabs\times \sum apec$), then a second fitting was done using typical WR abundances for the second absorption and for the emission component ($phabs\times vphabs\times \sum vapec$). These two abundance sets were chosen as they represent two extremes. On the one hand, the wind of the O-star companion is responsible for the whole emission; on the other hand, only the WR wind generates X-rays. As both winds are expected to contribute to the X-ray emission of a WR+O system, individually through embedded wind shocks or combined in a colliding wind zone, reality probably lies in-between the two considered extremes, but it is difficult (if not impossible) to know in advance the exact degree of mixing. The WR abundances were inspired by those assumed for Galactic models of PoWR\footnote{http://www.astro.physik.uni-potsdam.de/$\sim$wrh/PoWR/powrgrid1.php}. The WC and WNE PoWR models, adequate for our targets, consider the absence of hydrogen but Xspec uses abundances in number, with respect to hydrogen and with respect to solar values, hence hydrogen cannot formally be put to zero. Therefore, we decided to put the helium abundance to an arbitrary high value of $[n({\rm He})/n({\rm H})]_* / [n({\rm He})/n({\rm H})]_{\odot}=100$ and then scale the other element abundances relative to helium. This yields WNE abundances $[n({\rm X})/n({\rm H})]_* / [n({\rm X})/n({\rm H})]_{\odot}$ of 1, 550, 4.4, and 28 for C, N, O and other elements, respectively, and WC abundances $[n({\rm X})/n({\rm H})]_* / [n({\rm X})/n({\rm H})]_{\odot}$ of 7700, 65, 400, 56 for C, N, O and other elements, respectively. Fixing the He abundance to another value just scales the absorbing column and normalization factors according to the hydrogen abundance without any significant change on the other parameters (best-fit temperatures, fluxes) or on the fitting statistics ($\chi^2$). 

Final fitting results and corresponding fluxes are provided in Table \ref{fits}. Note that the results appear very similar for both abundance hypotheses, showing our results to be secure. In the Table, errors correspond to 1$\sigma$ uncertainties; whenever they were asymetric, the largest value is reported. Hardness ratios $HR$ were calculated as the ratios between the fluxes, corrected for interstellar absorption, in the hard (2.0--10.0 keV) and soft (0.5--2.0 keV) energy bands. Finally, the last column of Table \ref{fits} provides the luminosities corrected for interstellar and local absorptions (i.e. a proxy for intrinsic emissions) for completeness as some authors used them \citep[e.g.][]{zhe12}. However, their uncertainty must be underlined. The local absorbing column actually remains uncertain, in view of the existing trade-off between temperature and absorption: even for simple $phabs\times apec$ fittings, two solutions, one with high local absorption and low temperature and one with low local absorption and high temperature, provide fits of similar quality. The fluxes derived from these solutions, corrected by interstellar and local absorptions, may differ greatly, depending on the solution on which the fit converges - hence we considered such fluxes unreliable. This problem is particularly acute with short snapshots as ours but even exists with better data. As a concrete example, let us consider the various best-fits to high-quality data in \citet{naz08} and \citet{naz12} for the same (non-X-ray variable) star. The X-ray flux after correcting by both interstellar and local absorptions, vary between 0.9 and $7\times 10^{-11}$\,erg\,cm$^{-2}$\,s$^{-1}$, i.e. by one order of magnitude, depending on the choice of model. In our sample, the one-temperature fitting of WR\,127 yields a slightly worse fit quality than the two-temperature solution shown in Table \ref{fits}, but their X-ray luminosities after correction by the full absorbing column differ by a factor of 500. We therefore rather focus on the more reliable values of the luminosities corrected only for interstellar absorption, i.e. the X-rays emerging from the winds. In this context, note that \citet{owo99,owo13} demonstrated that the observed scaling relation $\log [L_{\rm X}/L_{\rm BOL}]\sim -7$ of OB stars can only be explained by considering the {\it emergent} X-ray luminosities and a delicate balance between emission and absorption by the wind: using emergent luminosities is therefore not unphysical when stellar winds are involved. 

We converted the count rates from faint detections or upper limits using WebPIMMS\footnote{https://heasarc.gsfc.nasa.gov/cgi-bin/Tools/w3pimms/w3pimms.pl}. We used an optically thin thermal model with solar abundances and temperatures of 0.6 and 6.9\,keV (a range covered by the spectra), with absorptions equal to the interstellar one. The derived range of ISM-absorption corrected luminosities are provided in Table \ref{conv}.

\begin{table}
\centering
  \scriptsize
\caption{Range of X-ray luminosities (in the 0.5--10.0\,keV energy band) derived from count rates and upper limits. }
\label{conv}
\setlength{\tabcolsep}{3.3pt}
\begin{tabular}{lccc}
  \hline\hline
Name & $N_{\rm H}^{\rm ISM}$ & $L_{\rm X}^{\rm ISM\,cor}$(tot, min-max) & $\log(L_{\rm X}^{\rm ISM\,cor}$(tot)\\
     & ($10^{22}$\,cm$^{-2}$) & ($10^{31}$\,erg\,s$^{-1}$) & $/L^{\rm WR}_{\rm BOL})$ - min,max\\
  \hline
\multicolumn{4}{l}{\it Upper limits on non-detections}\\
WR\,4   & 0.44 & $<$1.6-2.8 & $<$--8.1 to --7.8 \\
WR\,12  & 0.59 & $<$2.1-4.0 & $<$--8.3 to --8.0 \\
WR\,14  & 0.48 & $<$0.3-1.0 & $<$--8.2 to --7.9 \\ 
WR\,42  & 0.22 & $<$0.2-0.4 & $<$--9.0 to --8.6 \\
WR\,69  & 0.41 & $<$0.5-0.8 & $<$--8.2 to --8.0 \\ 
WR\,103 & 0.39 & $<$3.4-12. & $<$--8.5 to --8.0 \\
  \hline
\multicolumn{4}{l}{\it Faint detections}\\
WR\,9     & 0.83 & 13-22 & --7.1 to --6.8 \\
WR\,19$^a$& 0.98 & 16-36 & --6.6 to --6.2 \\
WR\,19$^b$& 0.98 & 15-31 & --6.6 to --6.3 \\
WR\,19$^c$& 0.98 & 23-39 & --6.4 to --6.2 \\
WR\,29    & 0.63 & 18-28 & --7.1 to --6.9 \\
WR\,30    & 0.46 & 3.4-11 & --7.8 to -7.3 \\
WR\,98    & 1.18 & 1.4-6.9& --8.1 to -7.4 \\
WR\,105   & 1.60 &  41-119& --6.9 to --6.4 \\
\hline      
\end{tabular}

{\scriptsize For WR\,14, the most stringent values (from ObsID 0780070801) are provided; for WR\,19, $^{a,b,c}$ correspond to ObsID 0792382601, 0793183301, and 0793183701, respectively.} 
\end{table}

\citet{pol87} and \citet{pol95} reported on detections of WR stars using {\it Einstein} and {\it ROSAT}, respectively. In \citet{pol87}, the log-likelihood of the detection threshold was fixed at 3: amongst our targets, WR\,105 and WR\,113 appeared to be near the threshold with IPC count rates of 0.008$\pm$0.007 and 0.002$\pm$0.003\,cts\,s$^{-1}$, respectively; WR\,97 lies above the threshold with an IPC count rate of 0.020$\pm$0.006\,cts\,s$^{-1}$. In \citet{pol95}, the log-likelihood of the detection threshold was fixed at 10 and WR\,97 appeared just at this threshold with a PSPC count rate of 0.029$\pm$0.012\,cts\,s$^{-1}$. In view of these values, all these detections were preliminary - indeed, they did not trigger further analyses. When converting the \xmm\ or {\it Swift} count rates of these targets, taking their spectral properties into account, IPC count rates of 0.036, 0.009, and 0.008\,cts\,s$^{-1}$ are found for WR\,97, 105, and 113, respectively, and a PSPC count rate of 0.04\,cts\,s$^{-1}$ is found for WR\,97. The reported and converted count rates agree relatively well, with probably a somewhat brighter emission for WR\,97 and 113 in \xmm\ data. 

\begin{figure*}
  \begin{center}
\includegraphics[width=5.8cm]{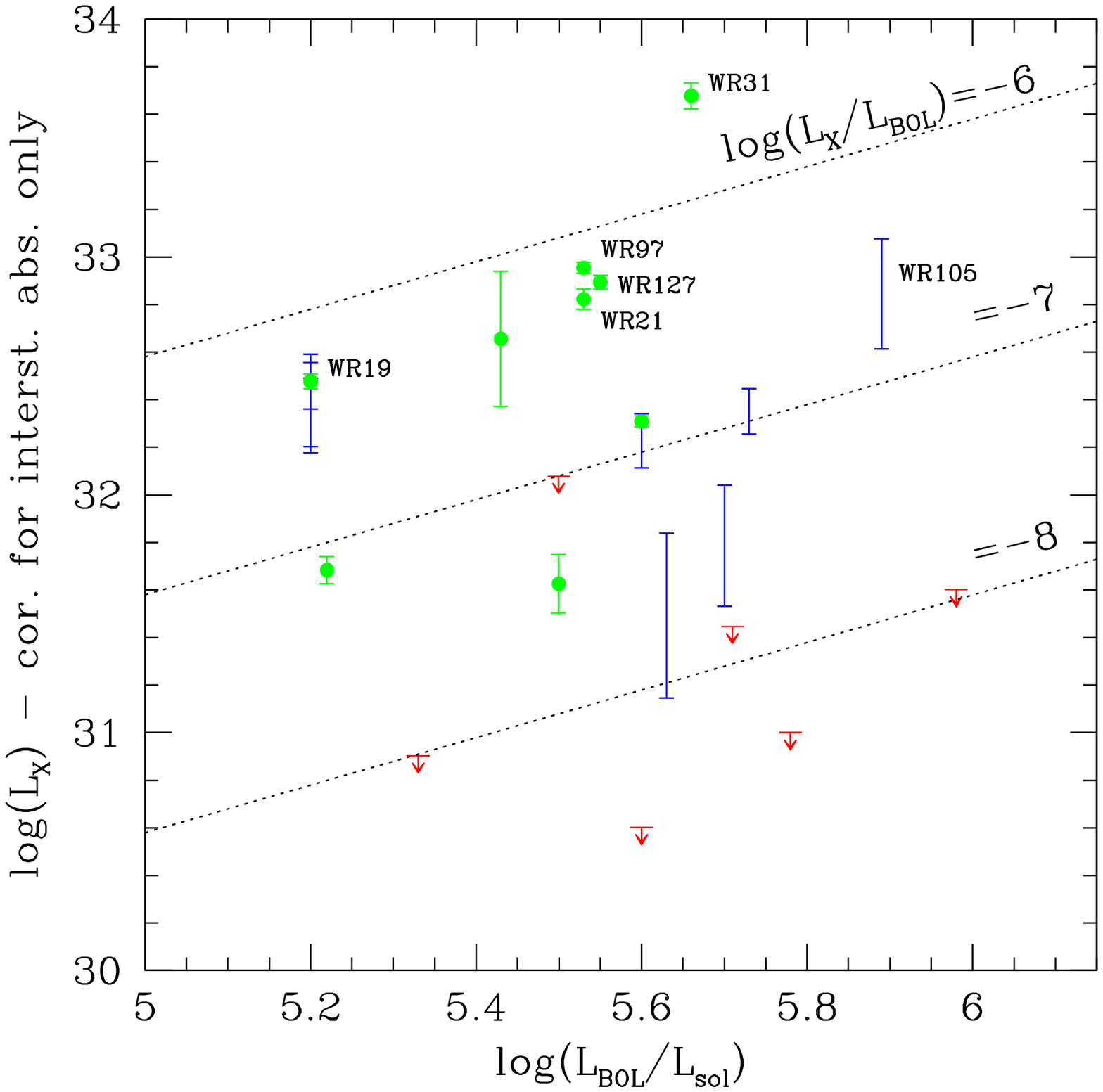}
\includegraphics[width=5.8cm]{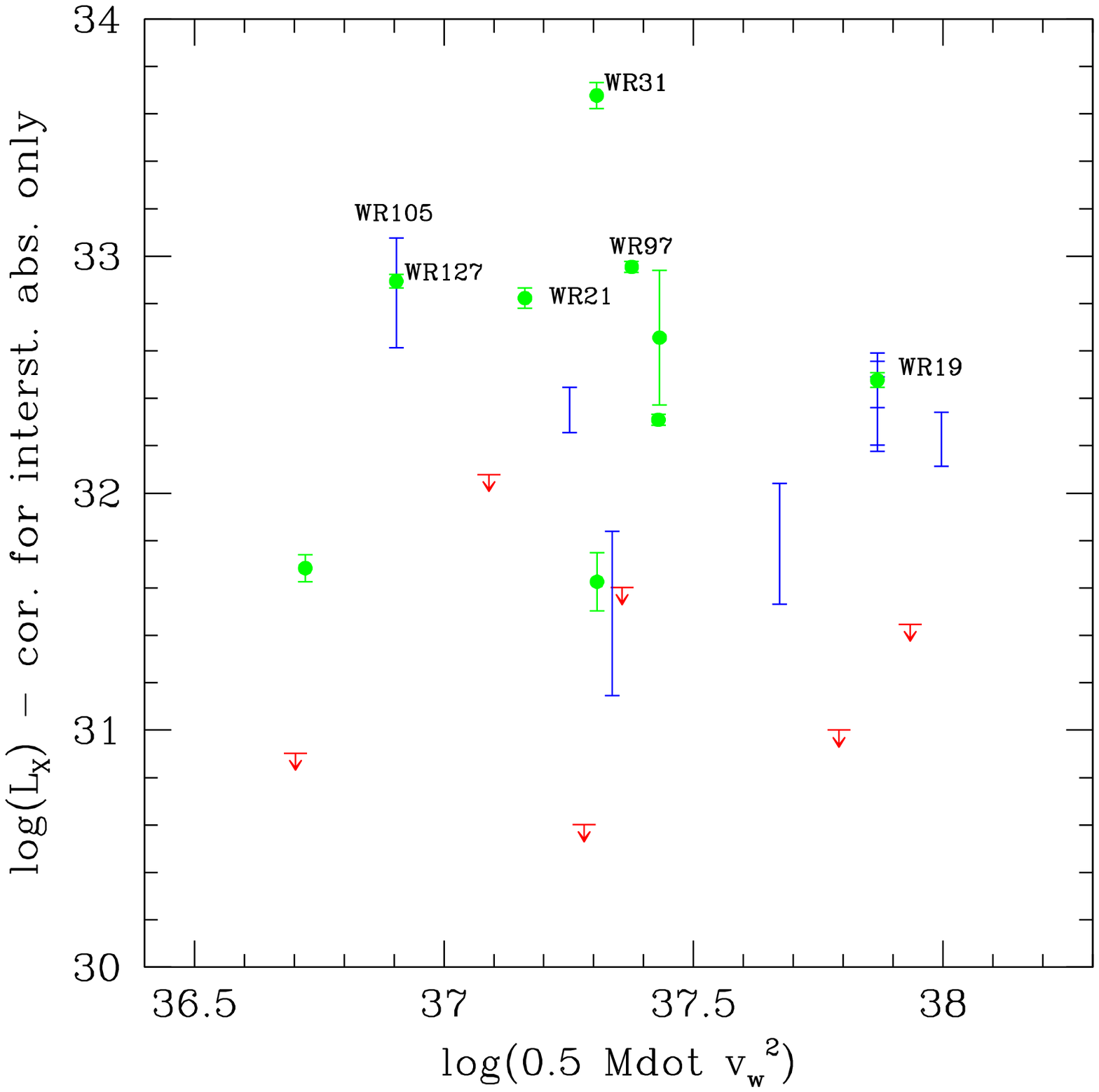}
\includegraphics[width=5.8cm]{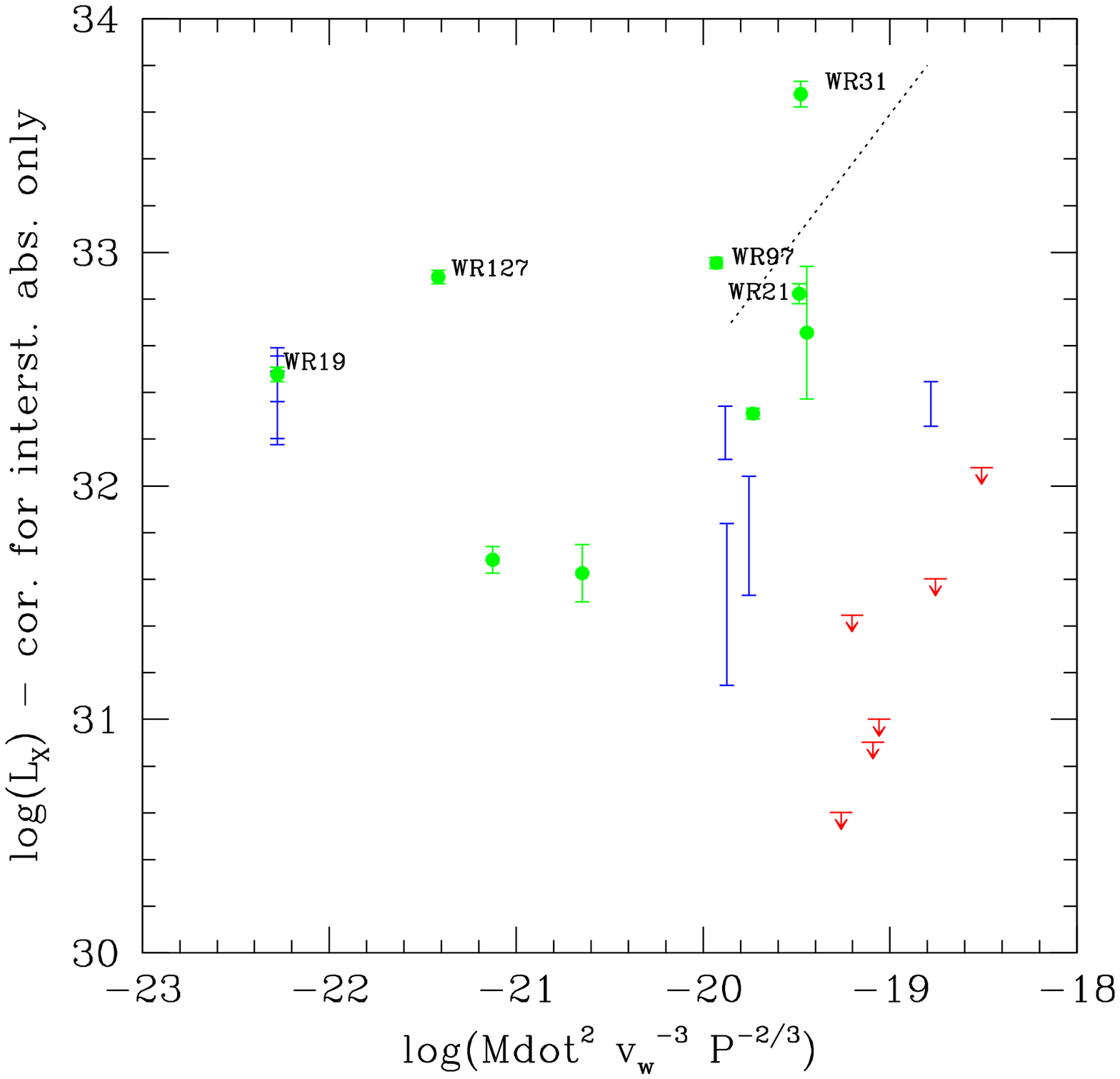}
\includegraphics[width=5.8cm]{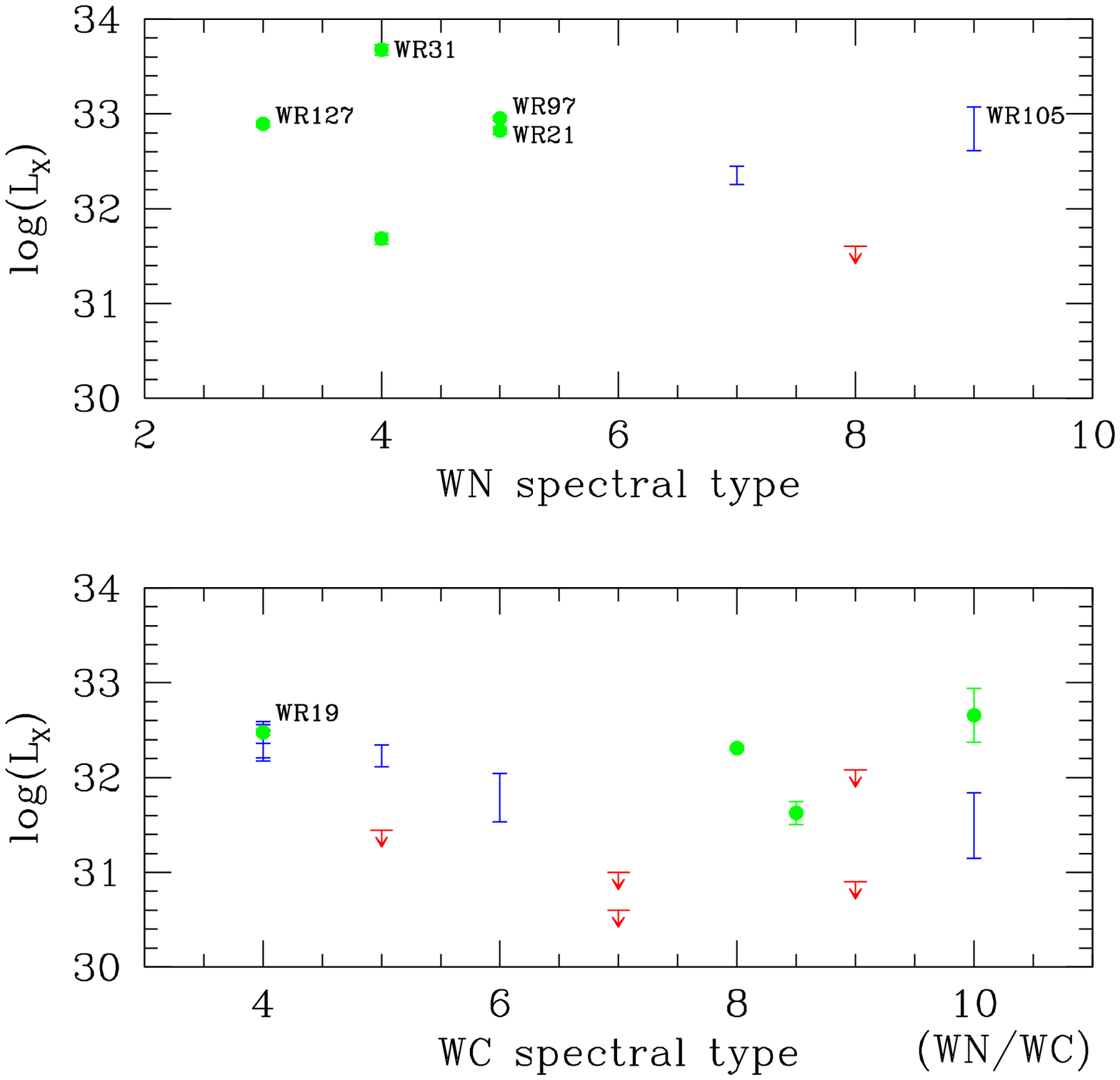}
\includegraphics[width=5.8cm]{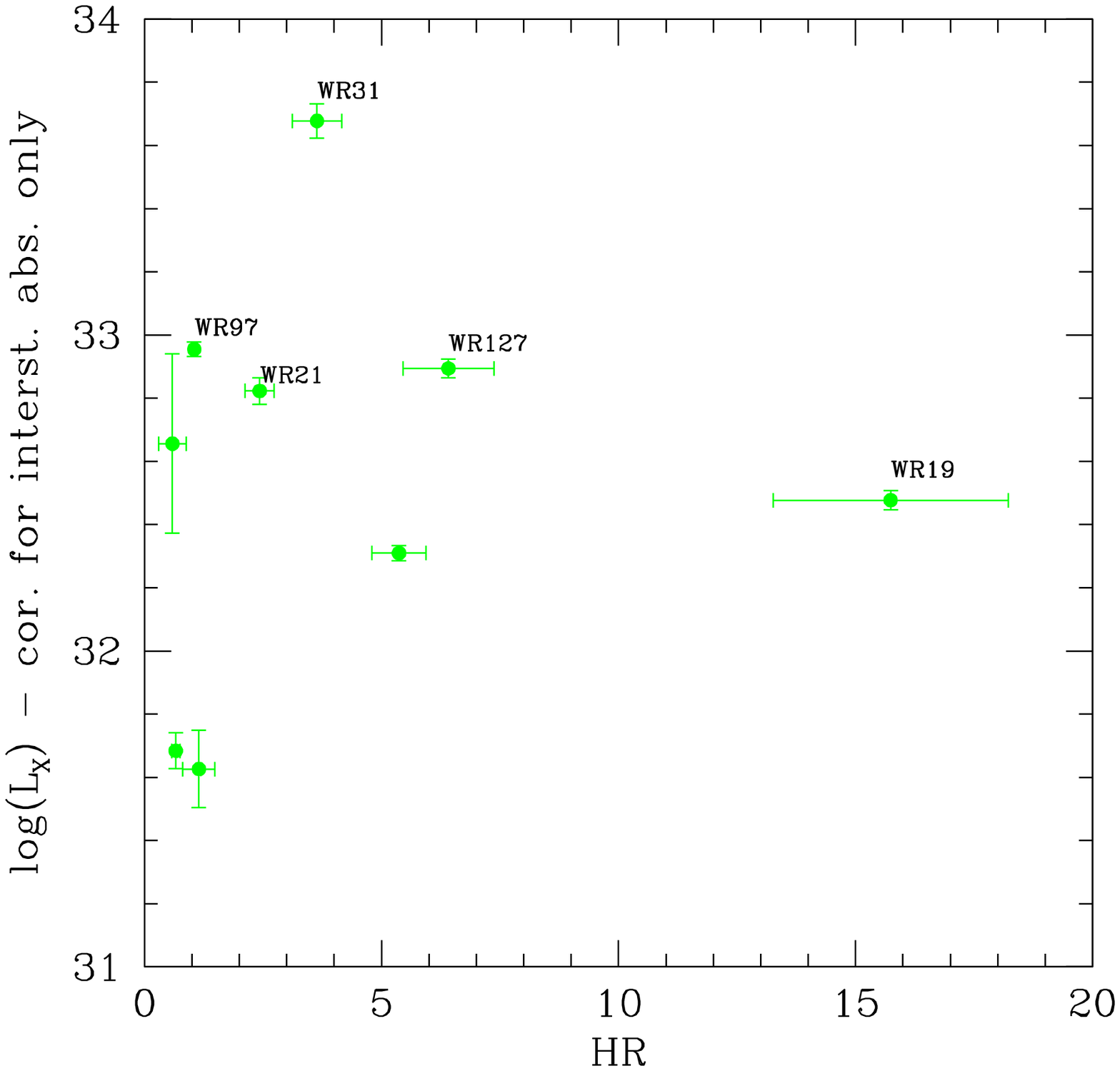}
\includegraphics[width=5.8cm]{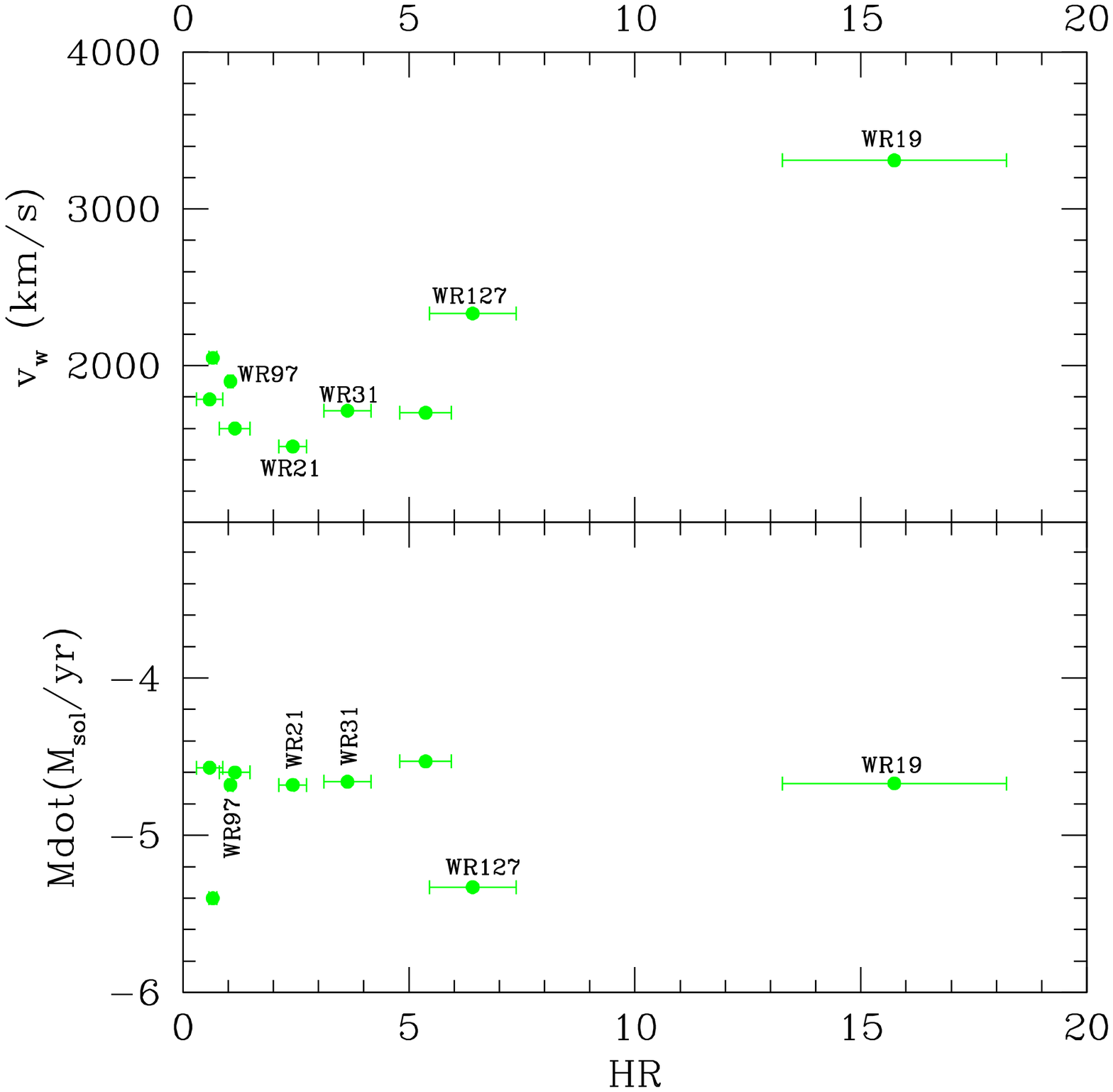}
  \end{center}
  \caption{Comparison of X-ray properties with the WR parameters. Upper limits are shown in red, faint detections in blue, and cases for which spectra were extracted in green (solar abundance fits from Table \ref{fits}). Hardness ratio corresponds to the ratio between ISM-absorption corrected fluxes in the hard (2.0--10.0\,keV) and soft (0.5--2.0\,keV) energy bands. }
\label{lx}
\end{figure*}

\section{Discussion}

These new X-ray snapshots provided information on 20 WR systems, but not all were detected in the X-ray range. While colliding-wind WR+O binaries are expected to display bright X-ray emission, six of our targets remain undetected. Their brightness limits are quite stringent, with $\log [L_{\rm X}/L^{\rm WR}_{\rm BOL}]<-8$ hence the permanent presence of a wind-wind collision appears unlikely. This is not especially surprising for WR\,14 and 69, as their $\sim$2\,d period came from photometry but without any subsequent confirmation of the presence of a companion, including in our photometric data. Furthermore, all but one undetected targets are WC stars, whose intrinsic X-ray emission is claimed to be very faint \citep{osk03}. However, this implies that some WC+OB systems appear X-ray faint, without significant contribution of colliding winds, while brightness seemed to be the rule for WC binaries up to now. The only non-WC star amongst the six non-detections is WR\,12, a WN+OB system for which X-ray observations were taken when the WR and its dense wind appear in front of the system (as seen from Earth, see phase in Table \ref{journal} and Fig. \ref{phot} for lightcurve). The non-detection may then be due to strong absorption at that phase (to realize the dramatic effect of wind absorption, see e.g. the case of WR\,22, \citealt{gos09}). An additional exposure, at another phase, would be needed before discarding the presence of colliding winds for this system.

Five other cases display a relatively faint but significant emission. Amongst these, only WR\,29 was observed with the WR in front, which probably explains the faintness of the X-rays. Two, WR\,9 and 30, are WC stars. As mentioned above, the intrinsic X-ray emissions of such stars are expected to be very weak hence the recorded X-rays could at first be thought as probably linked to colliding winds (but see below for an alternative explanation). An additional one, WR\,98, appears just below $\log [L_{\rm X}/L_{\rm BOL}]\sim -7$: it is difficult to assess how common this is for stars of this hybrid spectral type as only few of them were observed but the faintness of the emission probably argues against the presence of colliding winds (see also additional argument below). The last one, WR\,105, is of WN-type and appears brighter than expected from the typical relationship for OB stars ($\log [L_{\rm X}/L_{\rm BOL}]\sim -7$). This hints at the presence of X-ray emitting colliding winds, whose presence should be confirmed in more sensitive observations.

The nine remaining systems provided enough counts to obtain their spectral parameters. WR\,98a, 113 and 128 are the faintest cases but the temperature or hardness of their hot plasma appears somewhat high, even taking the usual temperature/absorption trade-off into account. Therefore, colliding winds could possibly contribute to their X-ray emission but more data (especially a monitoring along the orbit to search for orbital variations) need to be taken to assess that scenario. WR\,153ab appears moderately bright but also quite soft: additional data are here also required to assess the presence of a wind-wind collision (see also below). All other systems (WR\,19, 21, 31, 97, 127) have $\log [L_{\rm X}/L^{\rm WR}_{\rm BOL}]$ close to --6 and high temperatures or large hardness ratios, clearly pointing to wind-wind collisions. The hardest case is WR\,19 and this can be explained by a large local absorption, most probably due to the fact that this observation was taken at periastron passage, when the colliding winds are deeply embedded in the dense WR wind (see also Sect. 4.1) - again, a similar case as observed in WR\,22 \citep{gos09}. Other systems were not observed at phases with particularly large absorptions along the line-of-sights (see phases in Table \ref{journal} and phased optical lightcurves in Fig. \ref{phot}).

A last point needs to be examined before drawing conclusions. As the WR stars under study have O-type companions, the possibility should be considered that the X-rays are actually emitted not by the WR or a wind-wind collision, but by the O-star itself. We therefore compared the recorded X-ray luminosities to the typical bolometric luminosity of the companion (when its spectral type is more precisely known than ``O'' or ``OB'', see Table \ref{journal}) taken from \citet{mar05}. For WR\,42, the $\log [L_{\rm X}/L_{\rm BOL}]$ appears very low, for both WR and O bolometric luminosities. For WR\,30 and 98, the X-ray emission of the O-star with main sequence luminosity class could account for the X-ray emission. For WR\,9, the X-ray luminosity can be explained by the companion but only if its luminosity class was supergiant. For all other cases, the X-ray luminosities are too large for following the typical $\log [L_{\rm X}/L_{\rm BOL}]\sim-7$ of OB-stars, hence the alternative hypothesis of X-rays due to the sole companion can be rejected. However, for WR\,153ab, the X-ray emission level appears bright when compared to the bolometric luminosity of the sole WR component or of its companion, but if we combine the luminosities of all four components of this quadruple system, the $\log [L_{\rm X}/L_{\rm BOL}]$ reaches the typical value: combined to the softness of the recorded X-rays, this argues against the presence of X-ray bright colliding winds in the system. 

\begin{figure}
  \begin{center}
\includegraphics[width=8cm]{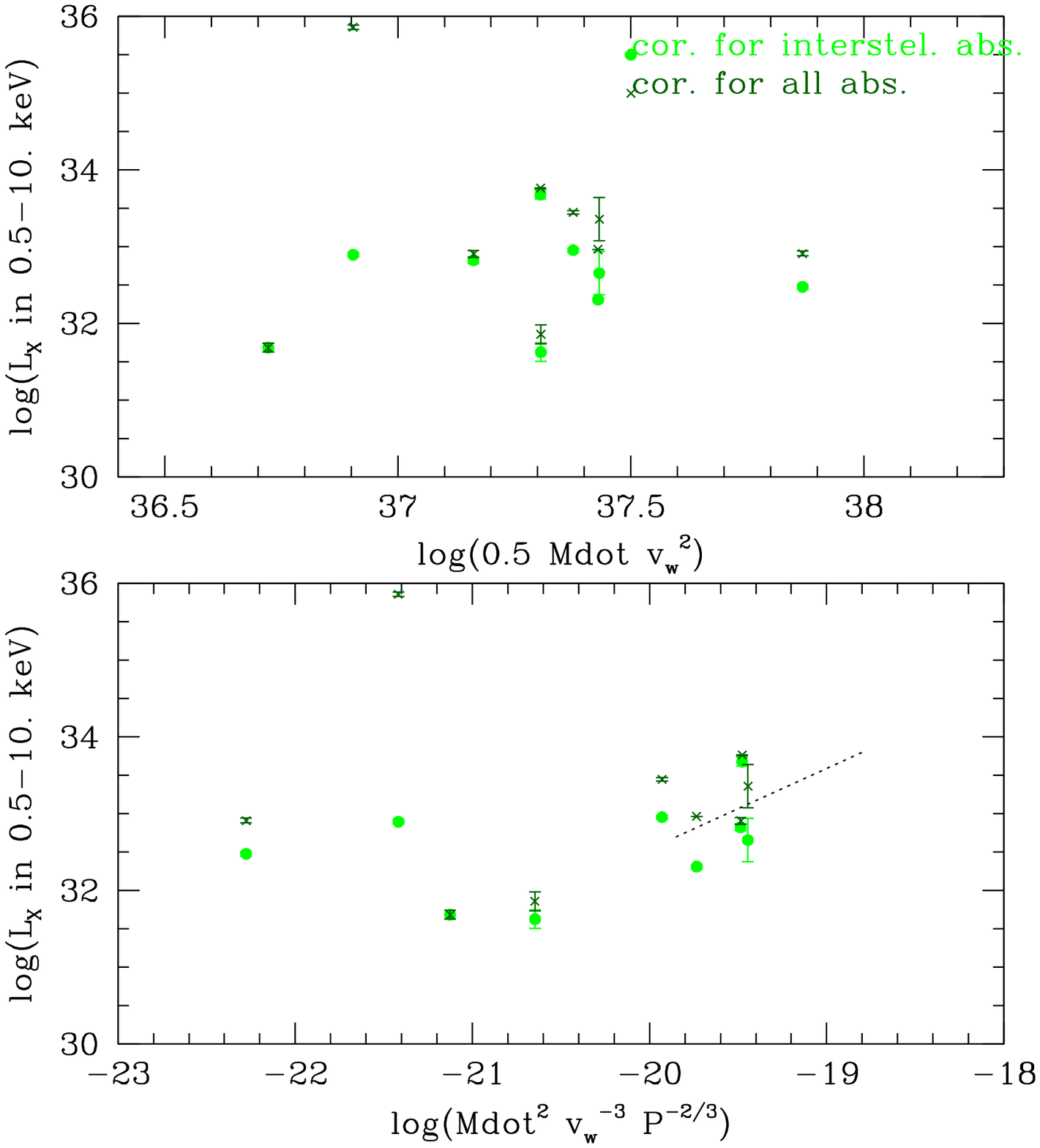}
\includegraphics[width=8cm]{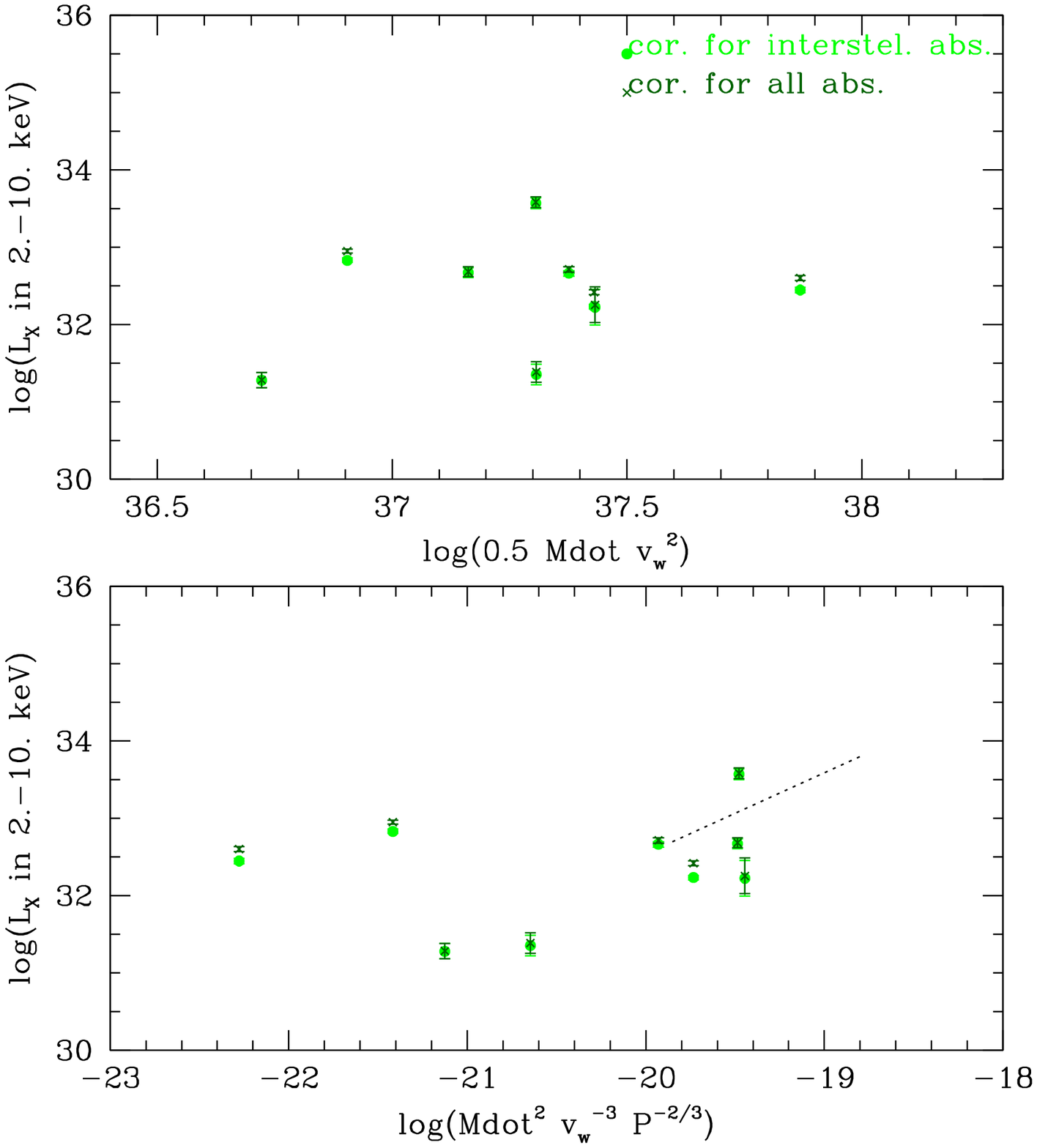}
  \end{center}
  \caption{Comparison of WR parameters and X-ray luminosities for several cases: luminosities corrected for interstellar absorption (light green symbols) or corrected for both interstellar and local absorptions (dark green symbols); total (0.5--10.\,keV, top) or hard (2.--10.\,keV, bottom) bands. }
\label{check}
\end{figure}

Figure \ref{lx} compares the observed X-ray luminosities and hardnesses to the known properties of the systems (Table \ref{journal}), i.e. WR spectral types, bolometric luminosities of the WR, WR wind velocities, mass-loss rates of the WR wind, WR wind luminosities ($=0.5 \dot M v_{\infty}^2$), and the parameter $\dot M({\rm M_{\odot} yr}^{-1})^2\times v_{\infty}({\rm km\,s}^{-1})^{-3} \times P({\rm d})^{-2/3}$. No typical value of the $\log [L_{\rm X}/L_{\rm BOL}]$ seems to exist for our targets, suggesting the absence of a single origin for X-rays in these stars. The hardness ratios also do not seem linked to the X-ray luminosity or wind parameters. For example, the mass-loss rate is quite low for WR\,127 and 128, but the local absorption is not particularly small (Table \ref{fits}); the wind velocity of WR\,19 is high but the plasma temperature does not correspond to the highest value in our fits (Table \ref{fits}). Furthermore, no obvious correlation is observed between wind luminosities and X-ray luminosities, and their ratio always appears very small ($<0.001$), both of which argues against a radiative nature for the wind-wind collisions, if present (\citealt{ste92}, although very low ratios may be possible in some specific radiative cases, see e.g. \citealt{kee14}). These low ratios confirm the results of \citet{zhe12}. On the other hand, no correlation either is found with $\dot M^2\times v_{\infty}^{-3} \times P^{-2/3}$, as expected for adiabatic collisions \citep{luo90,ste92} and reported by \citet{zhe12}\footnote{The $C_{\rm CSW}$ parameter of \citet{zhe12} uses another scaling as our $\dot M({\rm M_{\odot} yr}^{-1})^2\times v_{\infty}({\rm km\,s}^{-1})^{-3} \times P({\rm d})^{-2/3}$: to convert them, multiply the $C_{CSW}$ values by $10^{-19}$.}. To understand this difference, we need to compare the samples. Zhekov's targets were few in number (7) and some had quite large errors on their X-ray luminosities. On our side, our targets cover a larger range in $\dot M^2\times v_{\infty}^{-3} \times P^{-2/3}$, though only four systems display a parameter value lower than $10^{-20}$ (this is due either to long periods - WR\,19 and 98a - or to low mass-loss rates - WR\,127 and 128). But even focusing only on the same (more populated) interval as in \citet{zhe12} does not reveal a clear correlation between that parameter and the X-ray luminosities. \citet{zhe12} assumed that all recorded X-rays of his targets arose in colliding winds, but our sample shows that even WR binaries can appear faint. In fact, only six of our targets can be securely linked to colliding winds - they are identified in Fig. \ref{lx}. Focusing only on these more certain collision cases, we confirm the absence of correlation between hardness and wind velocity, mass-loss rate, or X-ray luminosity, as well as between X-ray and wind luminosities. While the X-ray luminosities of this subsample appear to increase with larger values of $\dot M^2\times v_{\infty}^{-3} \times P^{-2/3}$, the scatter around that relationship is large and its slope is shallower than derived by \citet[see dotted line in top right panel of Fig. \ref{lx}]{zhe12} hence it is difficult to confirm it. 

As a final note, we may underline that the conclusions of this paper (presence of non-detections or detections of systems with bright X-ray emissions, confirmation of the small fraction of wind luminosity converted into X-rays, absence of clear scaling of the X-ray luminosity with $\dot M^2\times v_{\infty}^{-3} \times P^{-2/3}$) remain whatever the choice is for the absorption correction (interstellar or interstellar+local) or if restricting the analysis to the hard band (which is largely unaffected by absorption effects) - see Fig. \ref{check}. This shows the robustness of our results. 

\subsection{Variation of the X-ray emission with orbital phase}
One defining characteristics for the presence of wind-wind collisions are their changes with orbital phase. In our sample, a few targets have been observed several times.

WR\,14 remained undetected in its two observations, despite very different orbital phases (Table \ref{journal}). However, the limit on the emission level only allows low X-ray luminosities (see above) hence the presence of X-ray bright colliding winds was already excluded.

WR\,105 was observed four times with {\it Swift}, without significant change of the count rate. As the orbital parameters of this system are not yet constrained, it is difficult to interpret this relative constancy in the context of colliding winds at present time.

The properties of WR\,97 in \xmm\ and {\it Swift} observations appear similar, showing that the X-ray emission has not changed much between them. However, these two exposures were done at similar orbital phases (Table \ref{journal}) hence this apparent constancy cannot be used to reject the presence of X-ray bright colliding winds in the system: a better coverage of the orbit would be needed to assess the presence of the expected phase-locked changes.

\begin{figure}
  \begin{center}
\includegraphics[width=8cm]{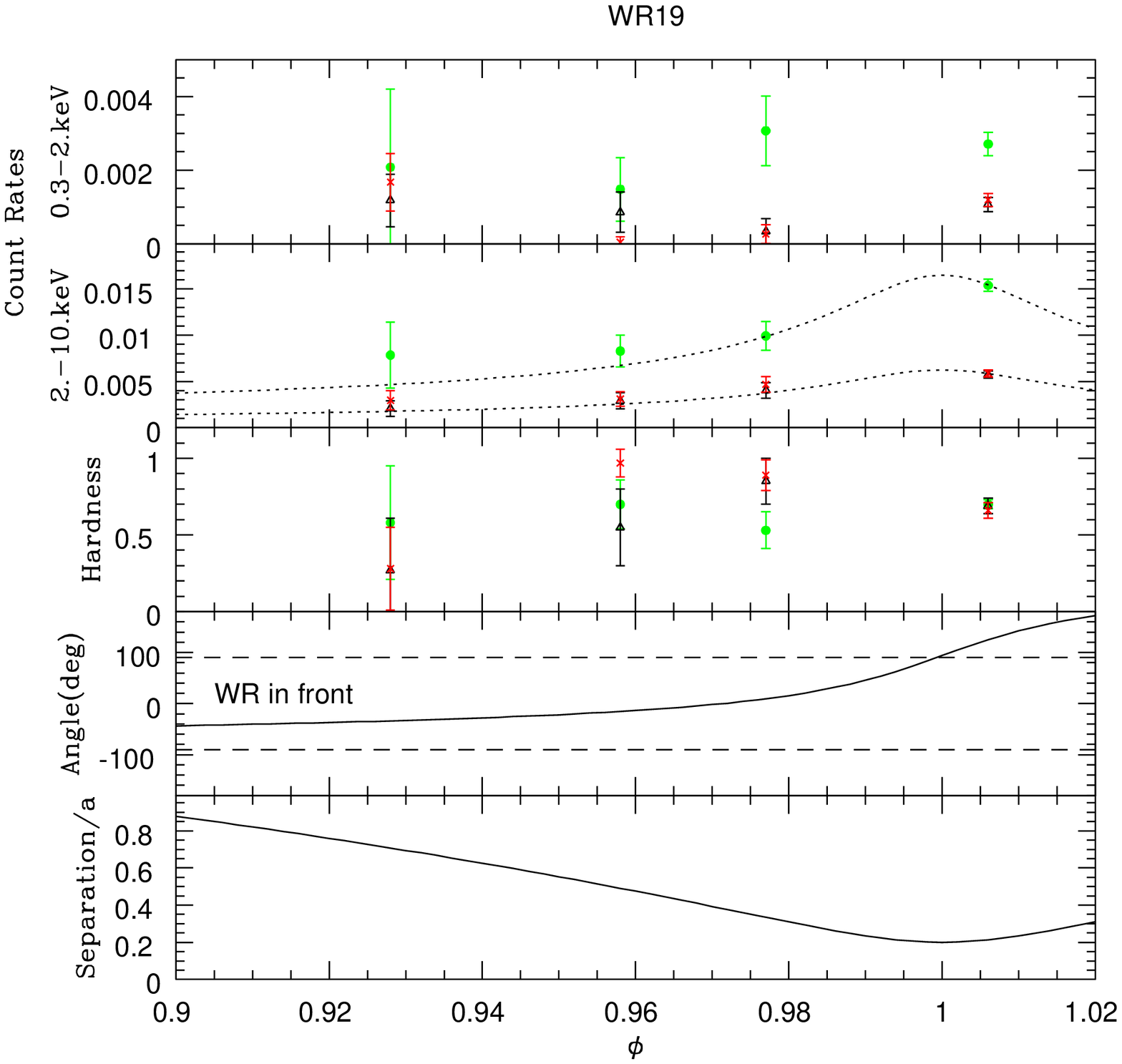}
\includegraphics[width=8cm]{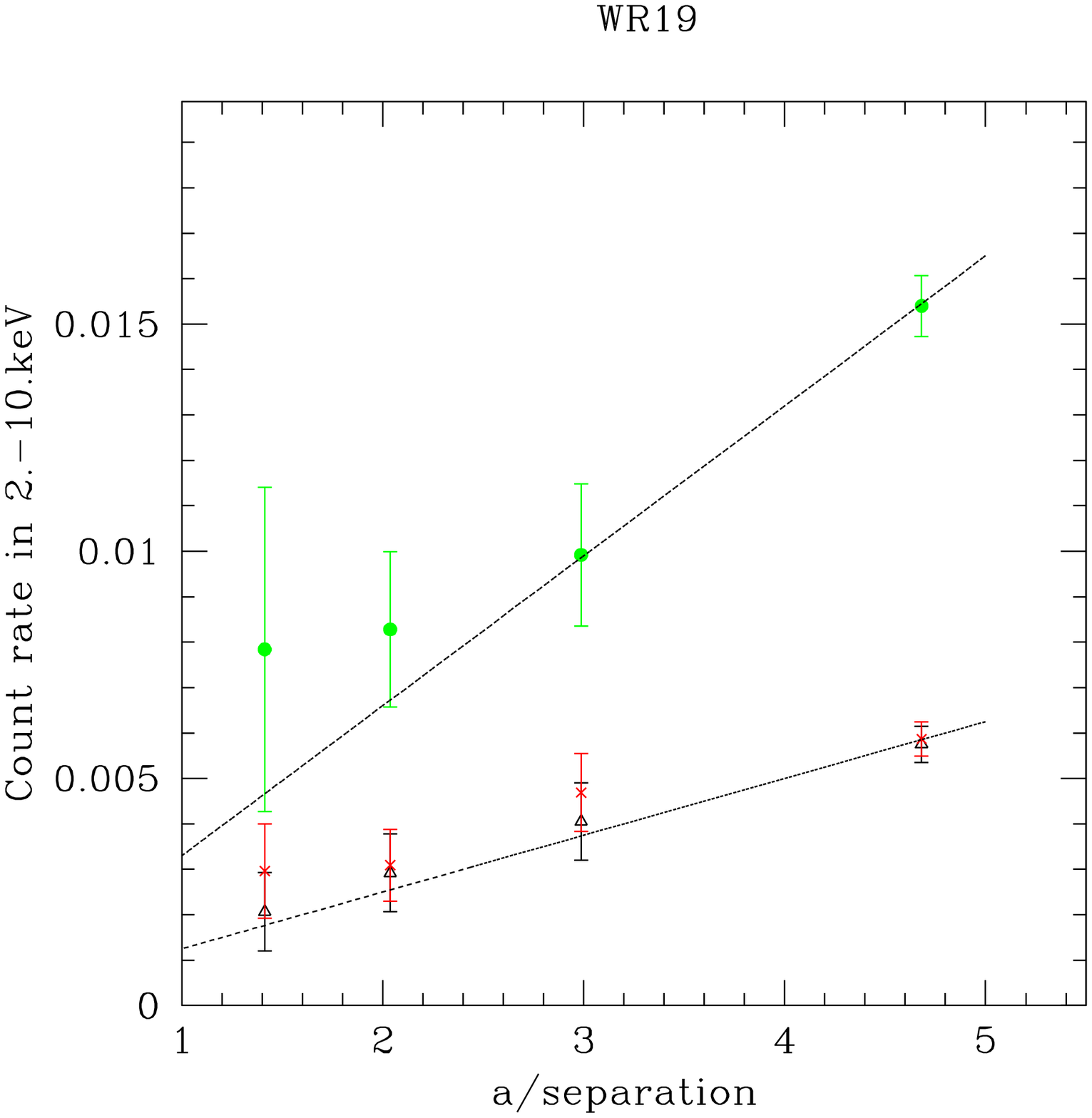}
  \end{center}
  \caption{Evolution of the EPIC (pn in green, MOS1 in black, MOS2 in red) count rates with phase in two energy bands, along with those of separation (normalized to the semi-major axis $a$) and orientation of the system (angle=0 when the WR star is in front, --90$^{\circ}$ and +90$^{\circ}$ at quadratures, marked by dashed lines) following the orbital solution of \citet{wil09}. Dotted lines display relationships of the type 1/separation. Hardness is here defined as $(h-s)/(h+s)$ where $s$ and $h$ are the count rates in the 0.3--2.0\,keV and 2.0--10.0\,keV energy bands.}
\label{wr19}
\end{figure}

Finally, WR\,19 was observed several times with {\it Swift} and \xmm. Two observations (ObsIDs 0792382601 for \xmm\ and 00034544005 for {\it Swift}) were taken ten days apart, i.e. at the same orbital phase since the period is long ($\sim$10\,yrs, Table \ref{journal}). Converting the \xmm\ count rates of this observation results into a {\it Swift} count rate of $\sim10^{-3}$\,cts\,s$^{-1}$ which corresponds to the recorded one (Table \ref{det}), demonstrating a good agreement between the two observatories. Since \xmm\ observations are more precise and better sample the periastron approach ($\phi_{\rm orb}$=0.9--1.0), we now focus only on these data. Overall count rates (Table \ref{det}) already revealed that the X-ray emission clearly appears to strengthen over time. To more precisely constrain these changes, spectra would be needed but the low number of counts (less than 100\,cts in EPIC-pn for non-periastron exposures) prohibits such an analysis. In replacement, an additional run of {\it edetect\_chain} provided the count rates in the 0.3--2.0\,keV and 2.0--10.0\,keV energy bands. Figure \ref{wr19} displays how these count rates evolve with orbital phase (using ephemerides from Table \ref{journal}). It compares them with the parallel changes in separation and orientation following the orbital solution of \citet{wil09}: $e=0.8, \omega_O=184^{\circ}$. Note the fast evolution of separation with phase near periastron, due to the very high eccentricity of the system.  The increase in the count rate of the harder X-ray band (which is largely unaffected by absorption or its changes) appears to closely follow the relationship with inverse separation expected for adiabatic wind-wind collisions \citep[see also preliminary hints in][]{sug17}. The X-ray emission of WR\,19 thus clearly presents all characteristics of a colliding wind emission. Finally, the hardness (defined as $(h-s)/(h+s)$ where $s$ and $h$ are the count rates in the 0.3--2.0\,keV and 2.0--10.0\,keV energy bands) increases towards periastron but the last point does not fit this trend, suggesting a drop in hardness although the error bars are quite large. Near periastron, due to closer separation, one expects a combination of stronger emission and large absorption (as found at that phase from spectral fitting, see Table \ref{fits}) but it is a priori difficult to know which process will win: clearly, further X-ray monitoring of the orbit with high quality, combined to a dedicated modelling, is now required to improve our understanding of this system. 

\section{Summary and conclusions}
This paper reports on dedicated X-ray snapshots of 20 WR+OB systems obtained with \xmm, {\it Chandra}, and {\it Swift}, with the aim of detecting new cases of X-ray bright wind-wind collisions. Six systems (WR\,4, 12, 14, 42, 69, 103) remain undetected and, except in one case (WR\,12) which could be explained by absorption from the WR wind, this faintness clearly is at odds with the presence of X-ray bright wind-wind collisions. It indicates for the first time that, as in O+OB binaries, not all WR+OB binaries are X-ray bright. Five other systems display a faint X-ray emission: in one case (WR\,29), this faintness could be due to the observation being taken as the dense WR wind appears in front hence could remain compatible with the presence of colliding winds; in three cases (WR\,9, 30, 98), the faint X-ray emissions could possibly be produced by the companion; one additional case (WR\,105) appears brighter than usually seen in massive stars hence is potentially compatible with the presence of colliding winds. For the last nine systems, enough counts could be recorded to constrain the spectral properties. Four of them (WR\,98a, 113, 128, 153ab) present X-ray to bolometric luminosity ratios compatible with the canonical relationship of massive OB stars, but five of them (WR\,19, 21, 31, 97, 127) display the bright and hard X-ray emissions typical of colliding winds. Moreover, one system (WR\,19) was monitored as the stars approach periastron and it presents the typical properties expected for adiabatic collisions: large absorption close to periastron (as the companion dives into the densest layers of the WR wind) and increase of emission as the separation decreases ($1/D$ relation). These detections represent a first step: further monitoring of the detected collisions are needed to constrain the collision properties in depth and allow for physical testing of these important phenomena.

\section*{Acknowledgements}
The authors acknowledge support from the Fonds National de la Recherche Scientifique (Belgium), the European Space Agency (ESA) and the Belgian Federal Science Policy Office (BELSPO) in the framework of the PRODEX Programme (contracts linked to \xmm\ and Gaia). ADS and CDS were used for preparing this document. 

\section*{Data availability}
The \te, {\it ASAS-SN}, \xmm, {\it Swift}, and {\it Chandra} data underlying this article are available in their respective archives.

\appendix

\section{Optical lightcurves}
Some of our targets are known to display photometric variations over the orbital period. As the ephemerides were often quite old, we tried to improve them using {\it TESS} \citep{ric15}, {\it Kepler} \citep{koc10}, and {\it ASAS-SN} \citep{sha14} data.

Two of the targets (WR\,4 and WR\,153ab) were observed with {\it TESS} at 2\,min cadence and the associated lightcurves were directly downloaded from the MAST archives\footnote{https://mast.stsci.edu/portal/Mashup/Clients/Mast/Portal.html}. Only the corrected lightcurves with the best quality (quality flag=0) data were considered. For the other stars, only \te\ full frame images (FFI) with 30\,min cadence are available. Individual lightcurves were then extracted for each target using the Python package Lightkurve\footnote{https://docs.lightkurve.org/}. Aperture photometry was done on image cutouts of 50$\times$50 pixels and using a source mask defined by pixels above a given flux threshold (generally 10 times the median absolute deviation above the median flux, but this was decreased if source is faint or increased if neighbours existed). Background (including scattered light) was then derived from the non-source pixels, using a principal component analysis (PCA with five components). Furthermore, the data points with errors larger than the mean of the errors plus their 1$\sigma$ dispersion were discarded. In addition, for WR\,97 and 98, the first hundred frames ($HJD<2\,458\,627$) were affected by a very local and intense patch of scattered light, hence they were discarded. Whatever the cadence, the raw fluxes were converted into magnitudes using $mag=-2.5\times \log(flux)+14$ (constant is arbitrary). In several cases, the targets were observed in two sectors and the lightcurves were then combined, after ensuring that the datasets shared the same mean. Both individual and combined lightcurves were analyzed. Note that three targets (WR\,98a, 105, and 113) do not have \te\ lightcurves.

{\it ASAS-SN} $V$ and $g$ lightcurves were extracted around the Simbad positions of the targets using the online tool\footnote{https://asas-sn.osu.edu/}. A few clearly discrepant points, i.e. points with magnitudes far from the rest of the dataset, were eliminated for WR\,4 (in V filter), WR\,21 (V), WR\,42 (both filters), and WR\,105 (g). 

A {\it Kepler} lightcurve is available for WR\,105 (=EPIC 224825377) and was downloaded from the MAST archives. Two consecutive observations of $\sim$20 and 40\,d exist, both the individual lightcurves and their combination were considered. The normalized fluxes were converted into magnitudes using $mag=-2.5\times \log({\rm norm.\ flux})$. Since normalized fluxes are used, the magnitude values are centered on zero by definition but a few clearly deviant points (with $mag > 0.05$) could be spotted and were discarded.

The \te\ and {\it Kepler} pixels have sizes of 21 and 4\arcsec, respectively, and the photometry is extracted from a few pixels while {\it ASAS-SN} images have a 15\arcsec\ FWHM PSF; hence crowding issues may arise. In Simbad, only WR\,103 has a bright neighbour within 1\arcmin\ radius. In Gaia-DR2, WR\,9, 19, 29, 30, 31, 98a, 127, and 128 have bright neighbours, while other stars have no neighbour with $\Delta(G)<2.5$ within 1\arcmin\ radius. Caution must thus apply to the former 8 stars.

All lightcurves were then analyzed using a modified Fourier algorithm adapted to uneven samples \citep{hmm,gos01,zec09}, as there is a small hole in the middle of each sector lightcurve. In half of the cases, a peak was clearly seen in the periodogram at the known orbital period. The optical data were then used to improve the ephemerides, in particular to derive an orbital reference time $T_0$ closer to today (Table \ref{journal}). Usually, the previous period determination was more precise than derived from our data, hence it was kept, except for WR\,30, 42, and 127. Table \ref{journal} provides the final ephemerides, as well as the phase of the X-ray observations, while Figure \ref{phot} shows the optical lightcurves folded with them. For the second half of the sample, no peak is present in the periodogram at the previously reported period. In WR\,4, 9, 14, 97, 113, and 128, the optical lightcurve mostly consists of noise, so we cannot confirm photometrically the claimed periods. On the other hand, \te\ does not sample well the long periods of WR\,19, 98, and 98a, as does {\it ASAS-SN} for WR\,19, and the {\it ASAS-SN} data of WR\,98 and 98a do not show any significant variation on the reported period. We may nevertheless note that a photometric minimum recorded during \te\ observations of WR\,98 occurs close to the expected conjunction (with the WR star in front). In parallel, the periodograms of WR\,69 and 103 reveal timescales different from previous reports, but they are not single: they are 0.207, 0.253, and 0.379\,d$^{-1}$ for WR\,69, and 0.200 and 0.780\,d$^{-1}$ for WR\,103 - the origins of these signals remain to be determined. Finally, no period is available for WR\,105 and it shows no clear long-term trend in {\it ASAS-SN} data, but the presence of short-term variations is clear from {\it Kepler} data (Fig. \ref{phot2}).

\begin{figure*}
  \begin{center}
\includegraphics[width=5.8cm]{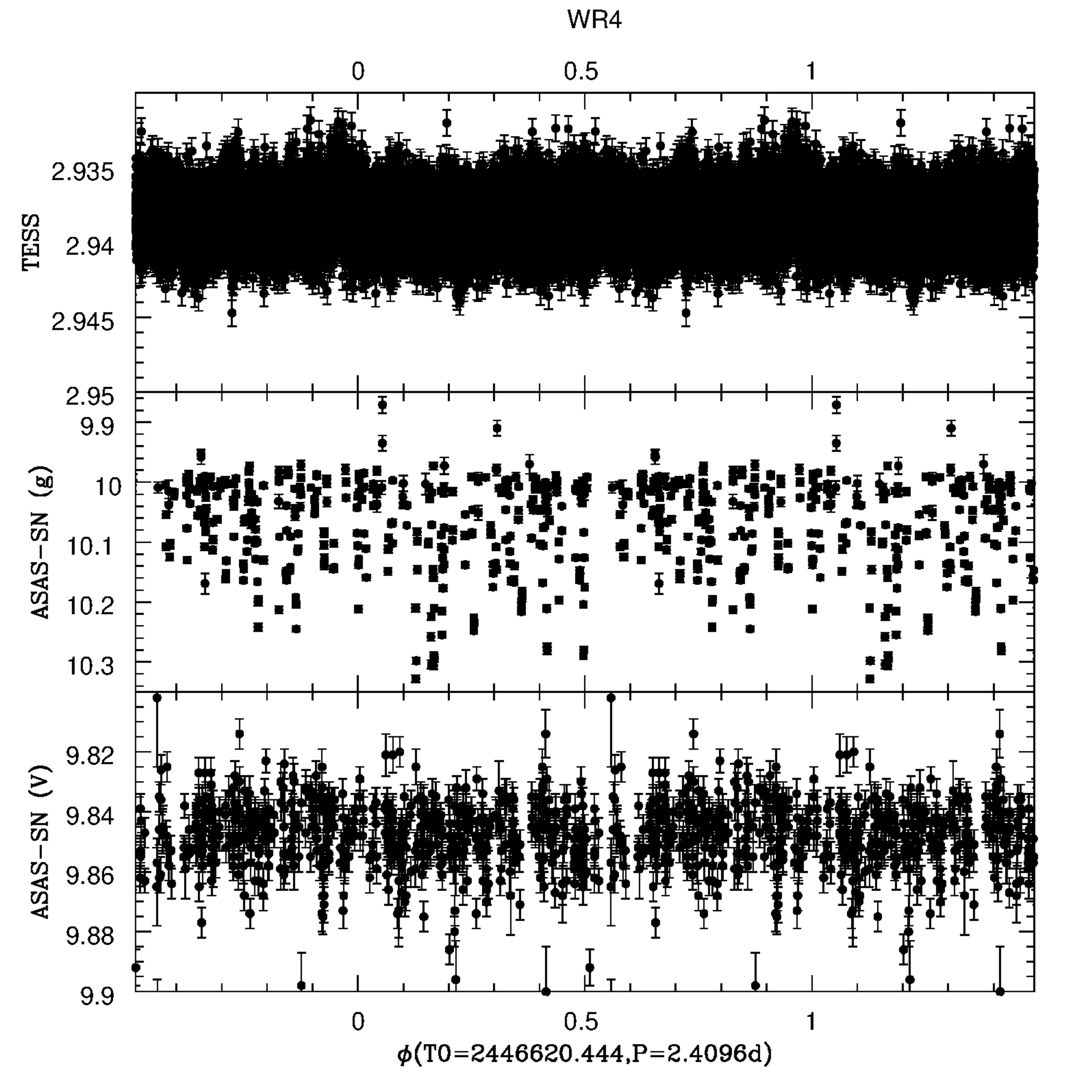}
\includegraphics[width=5.8cm]{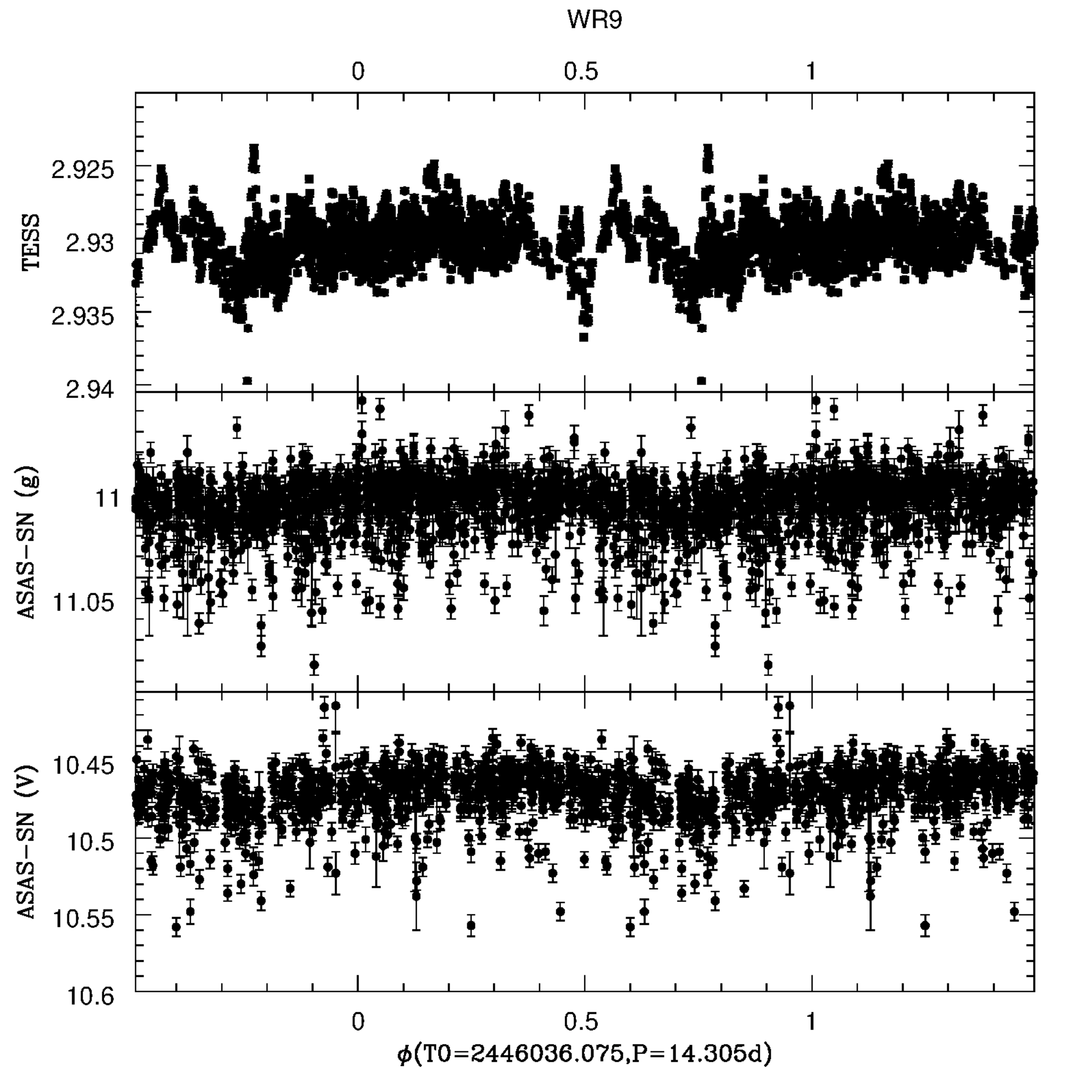}
\includegraphics[width=5.8cm]{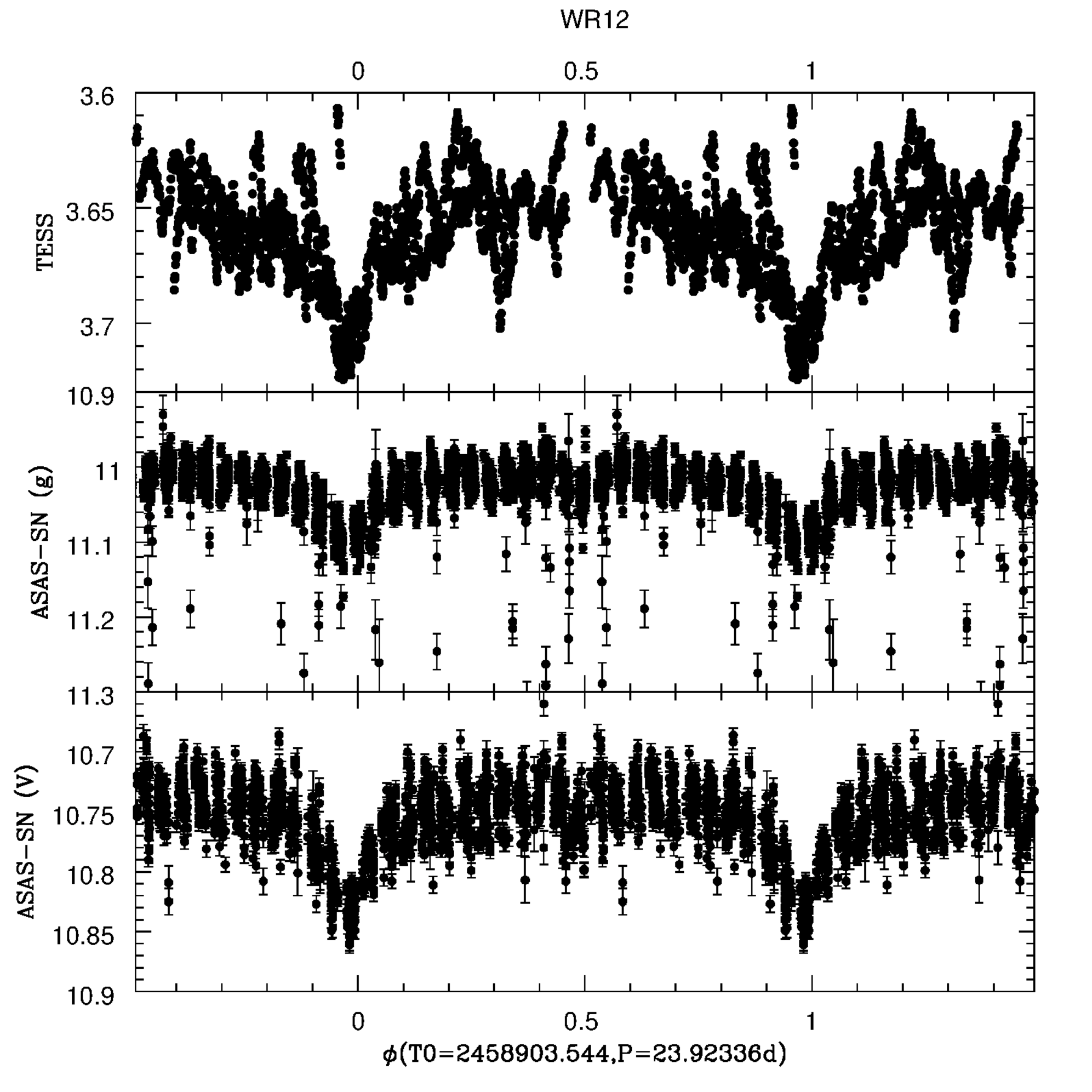}
\includegraphics[width=5.8cm]{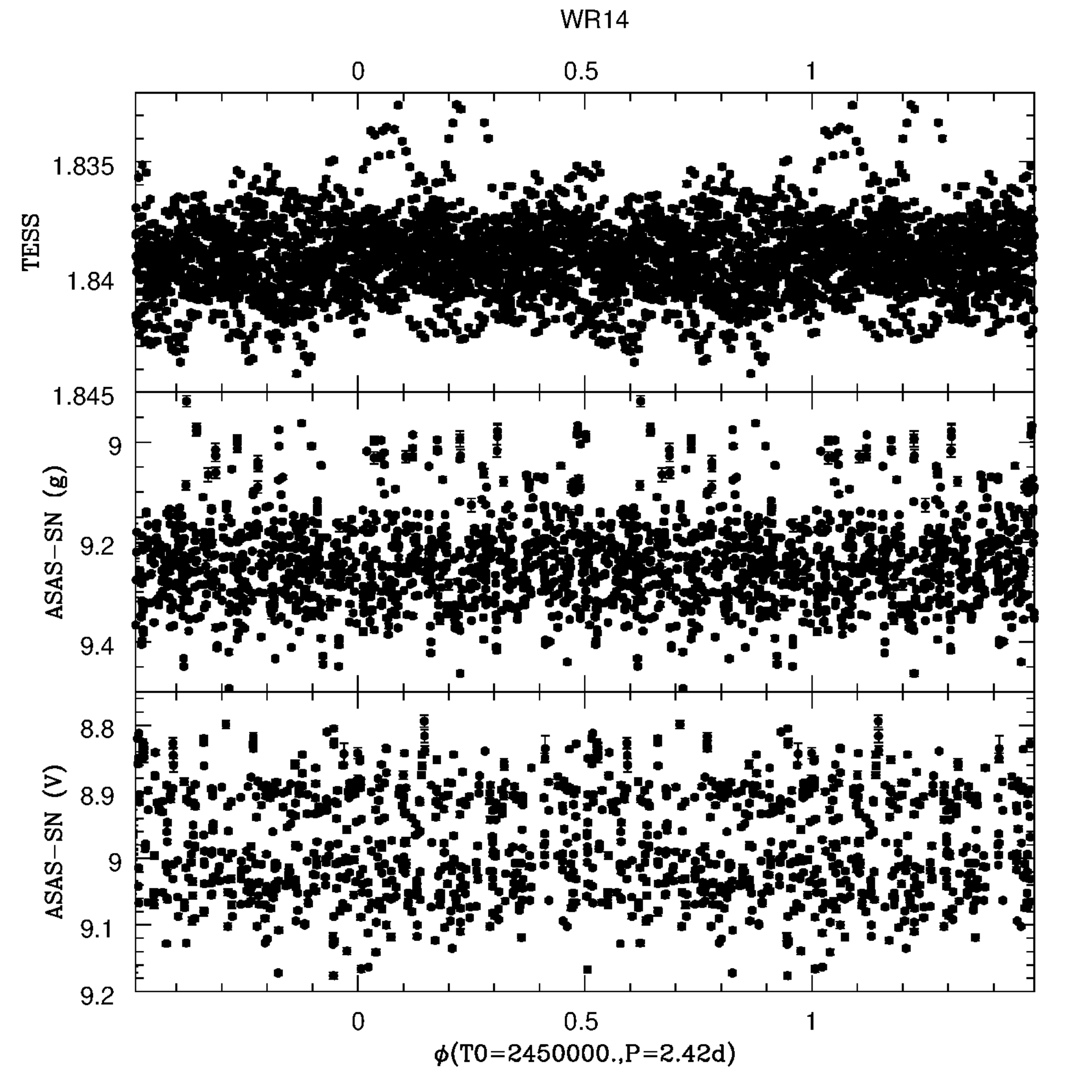}
\includegraphics[width=5.8cm]{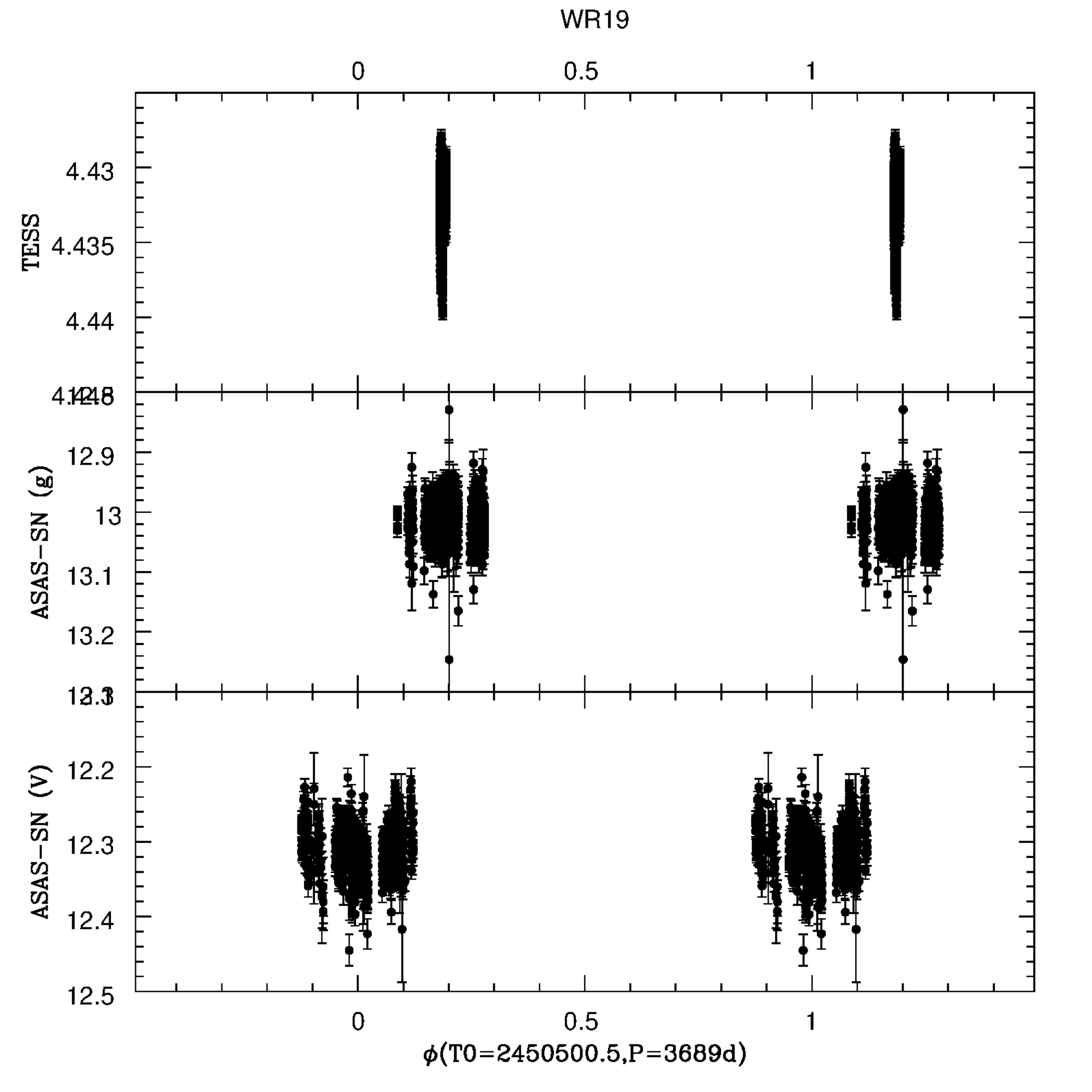}
\includegraphics[width=5.8cm]{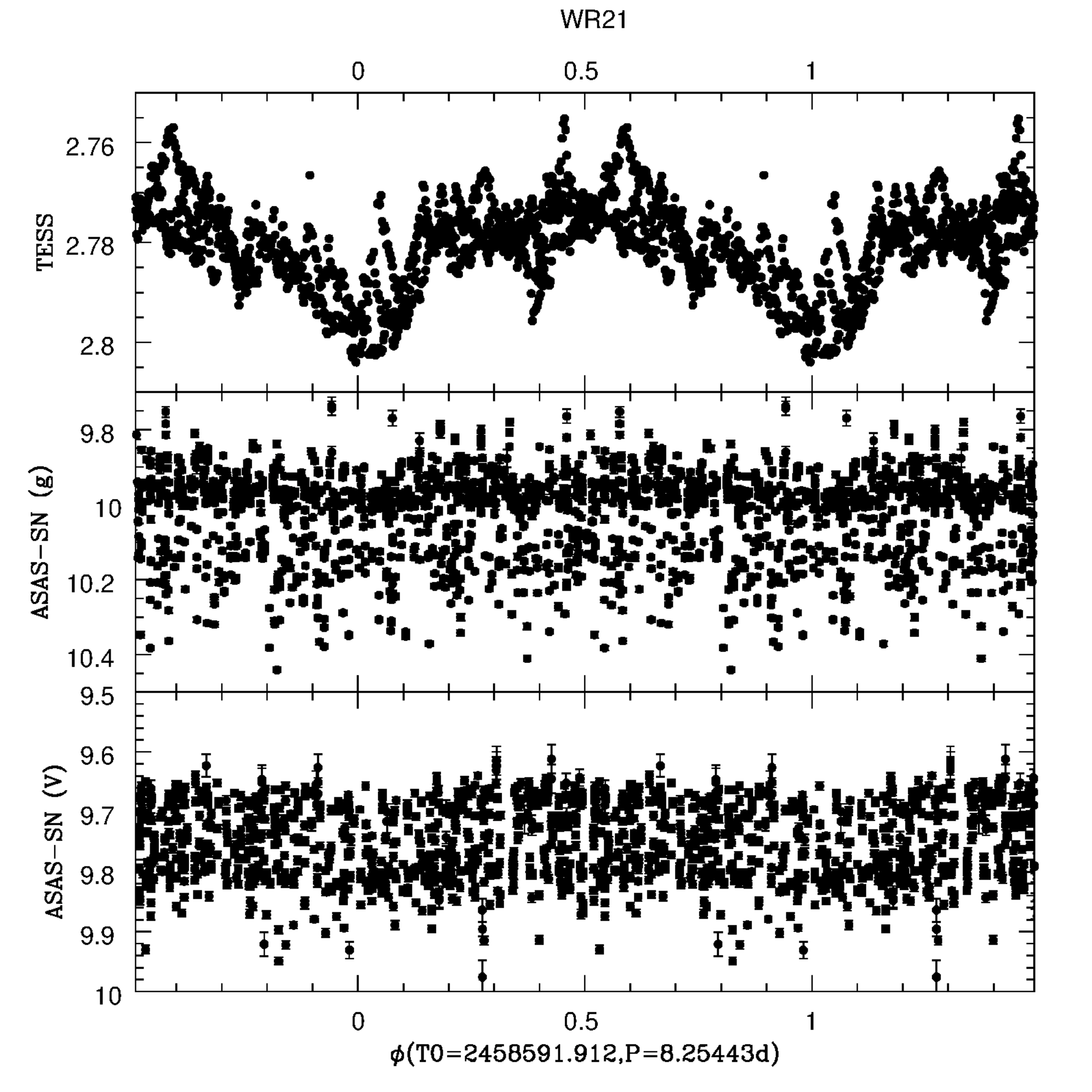}
\includegraphics[width=5.8cm]{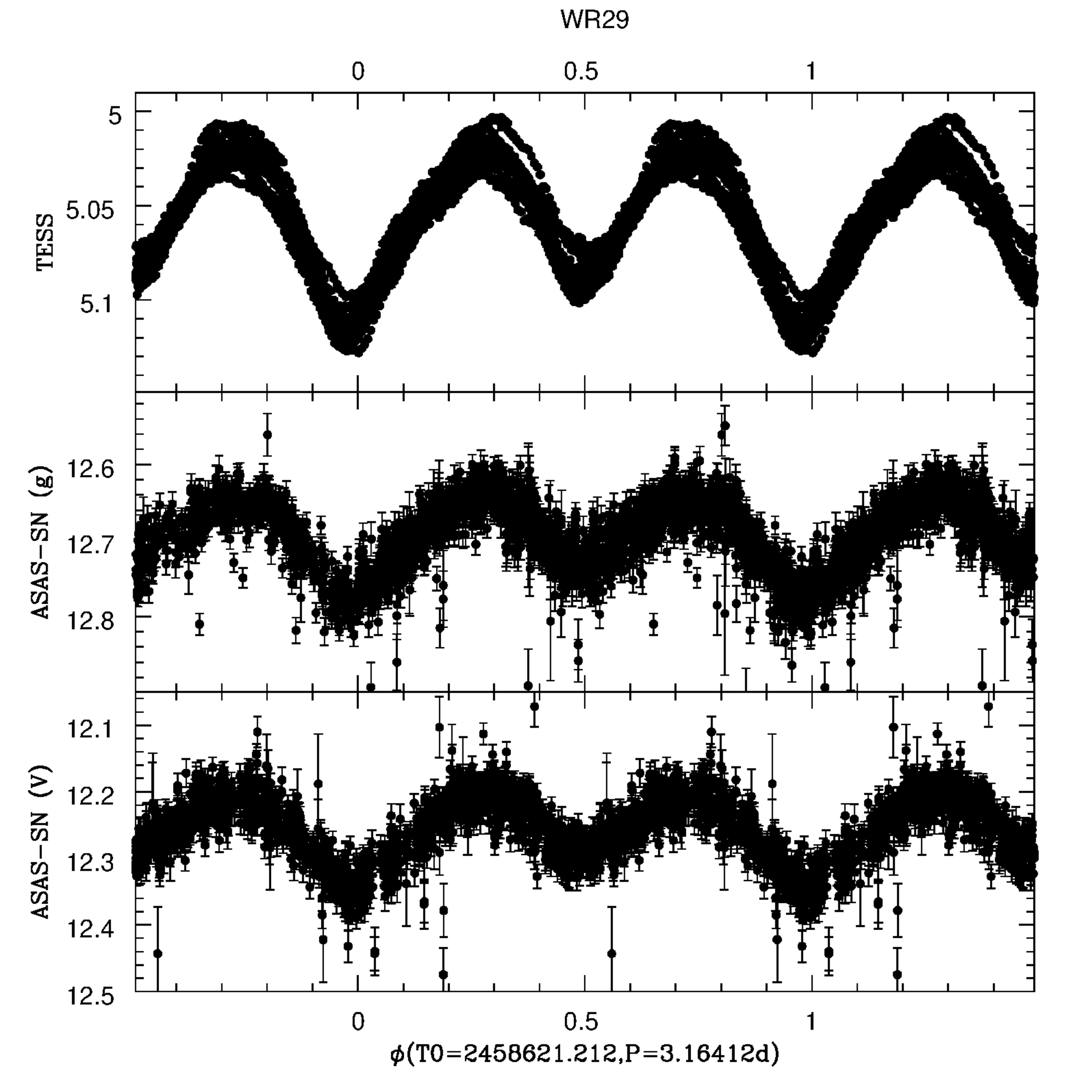}
\includegraphics[width=5.8cm]{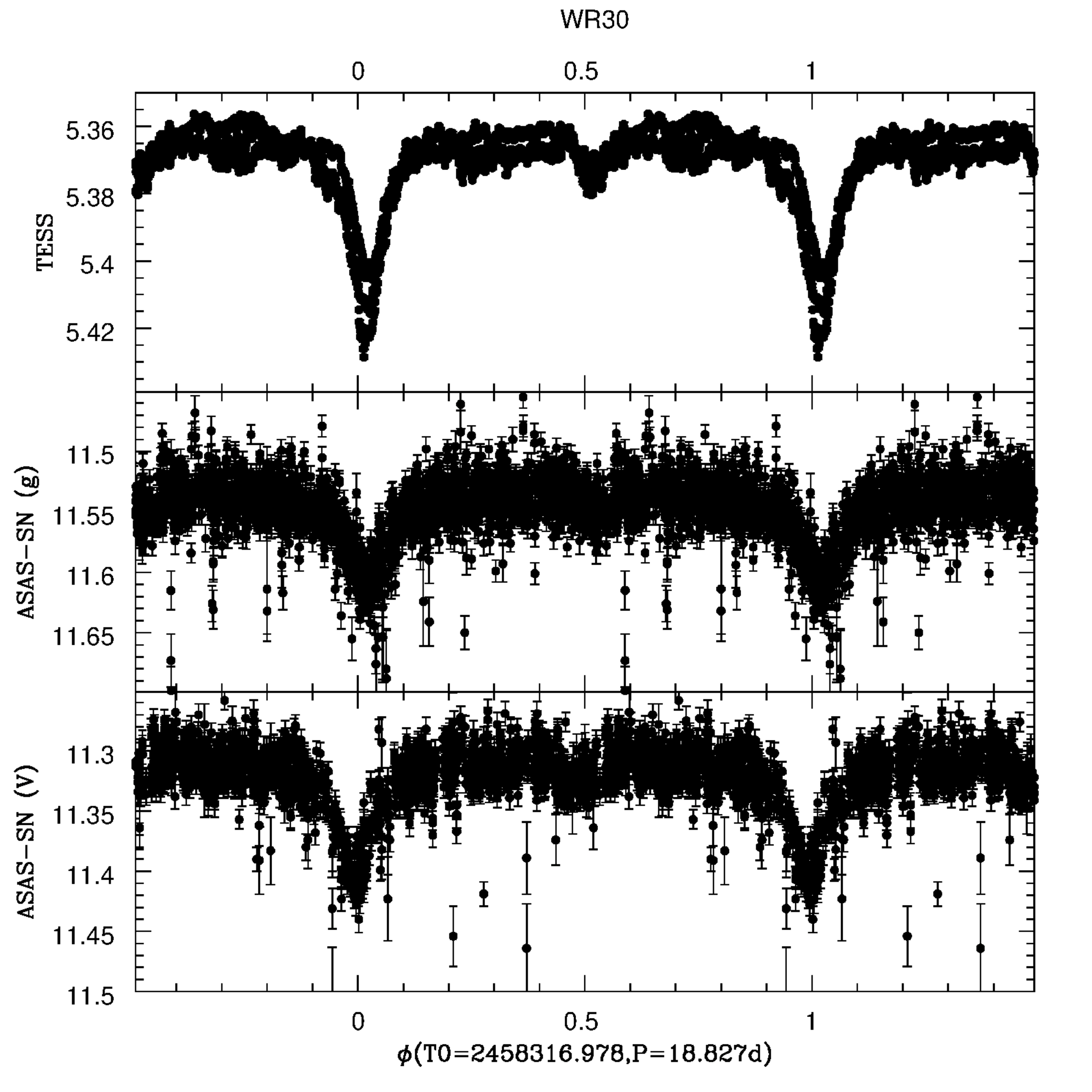}
\includegraphics[width=5.8cm]{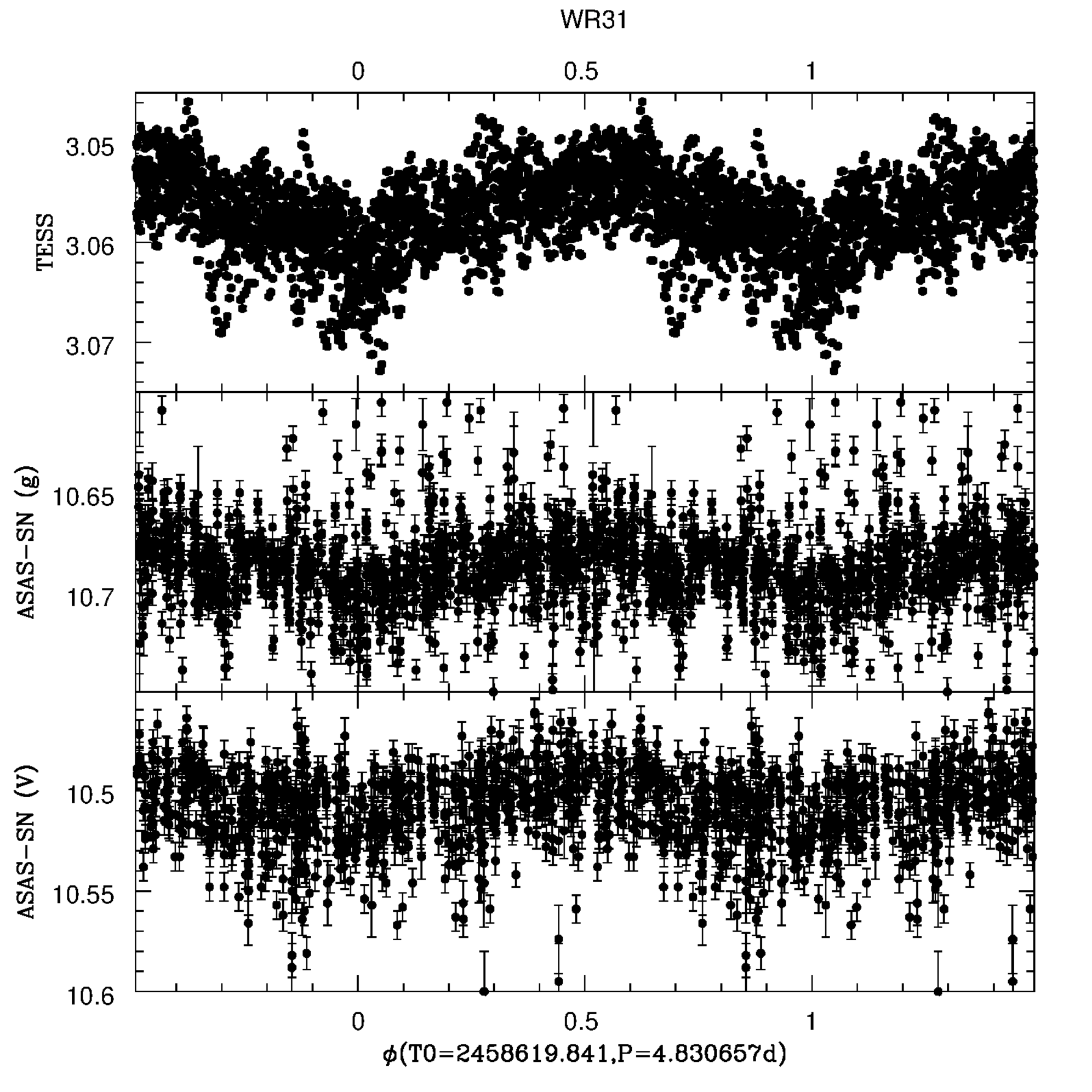}
\includegraphics[width=5.8cm]{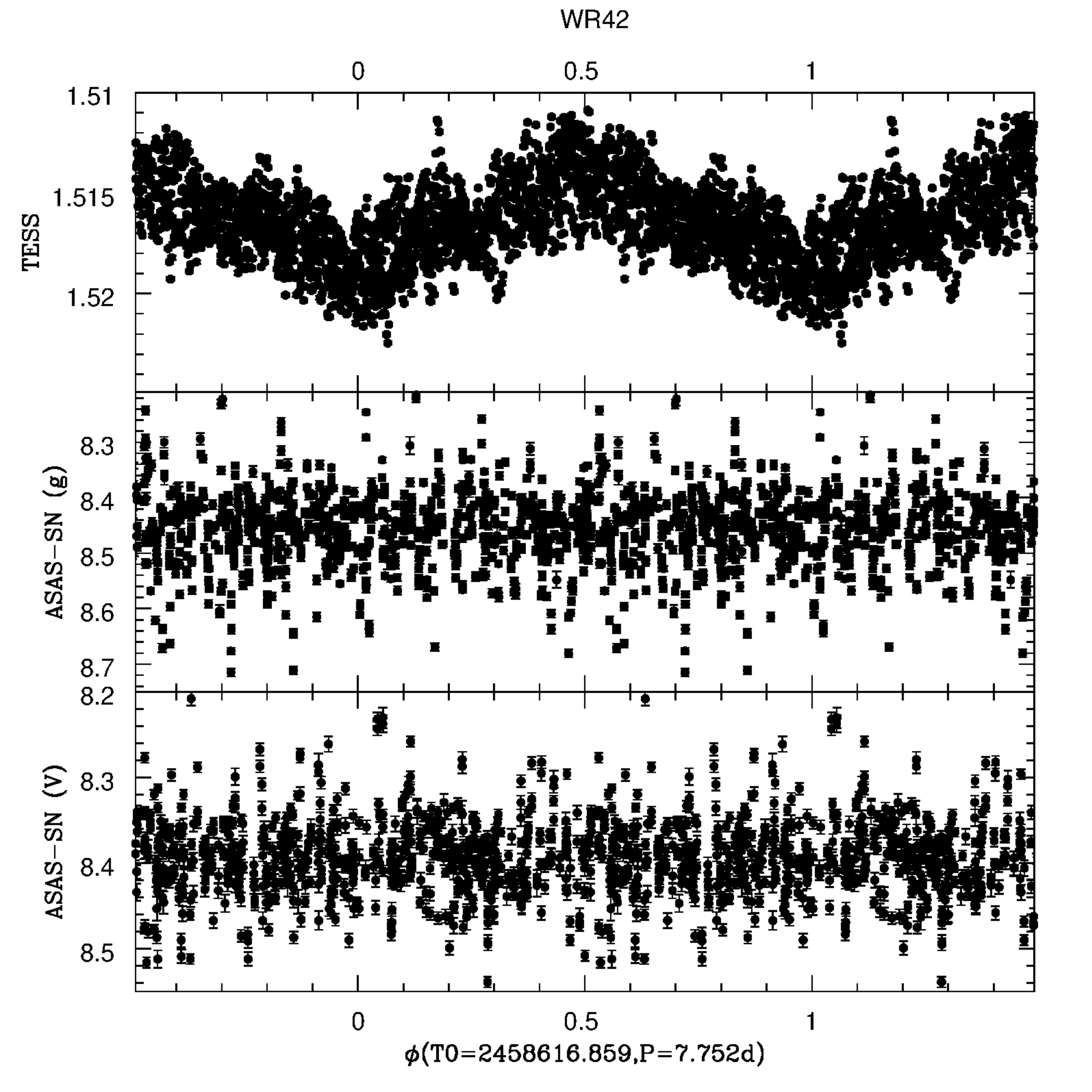}
\includegraphics[width=5.8cm]{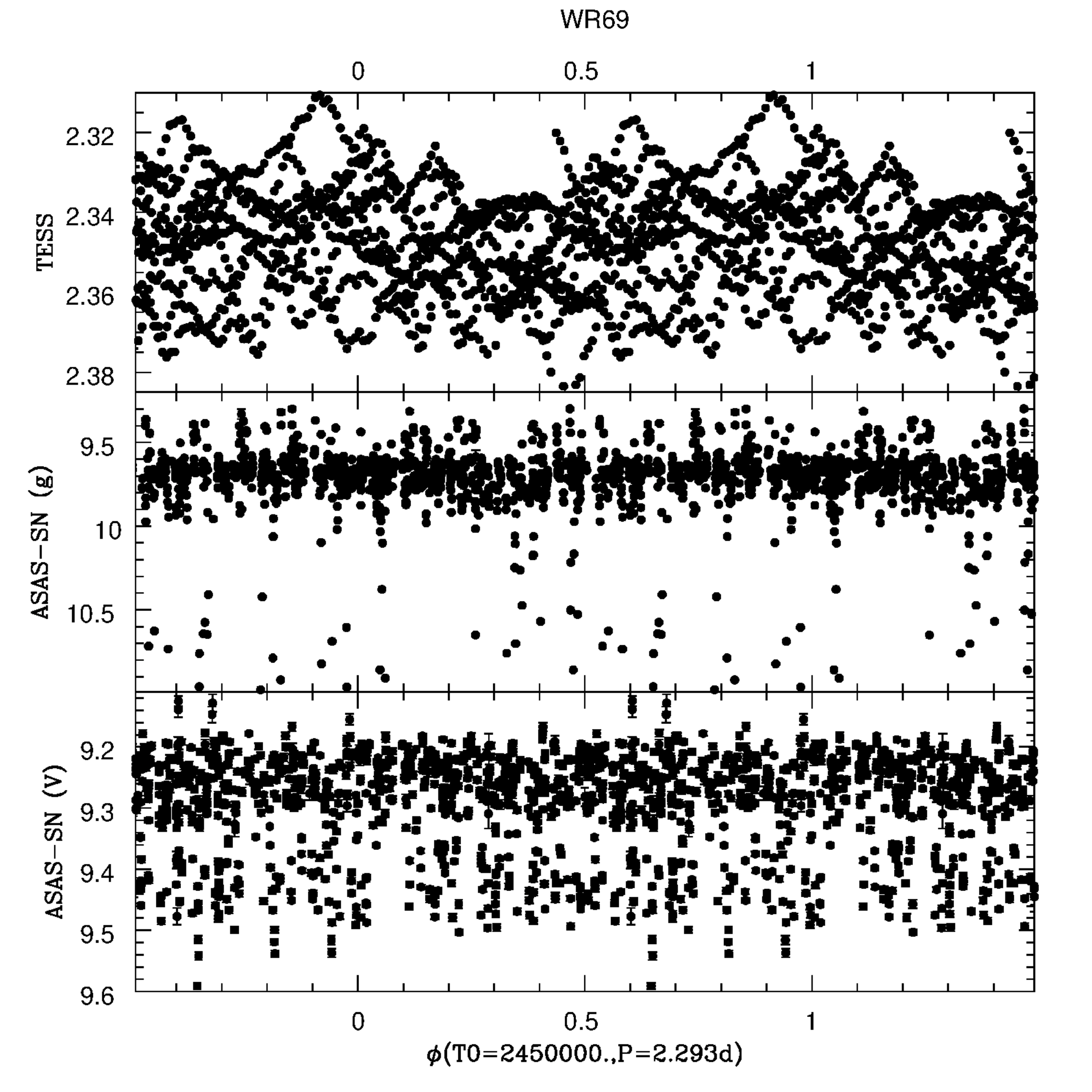}
\includegraphics[width=5.8cm]{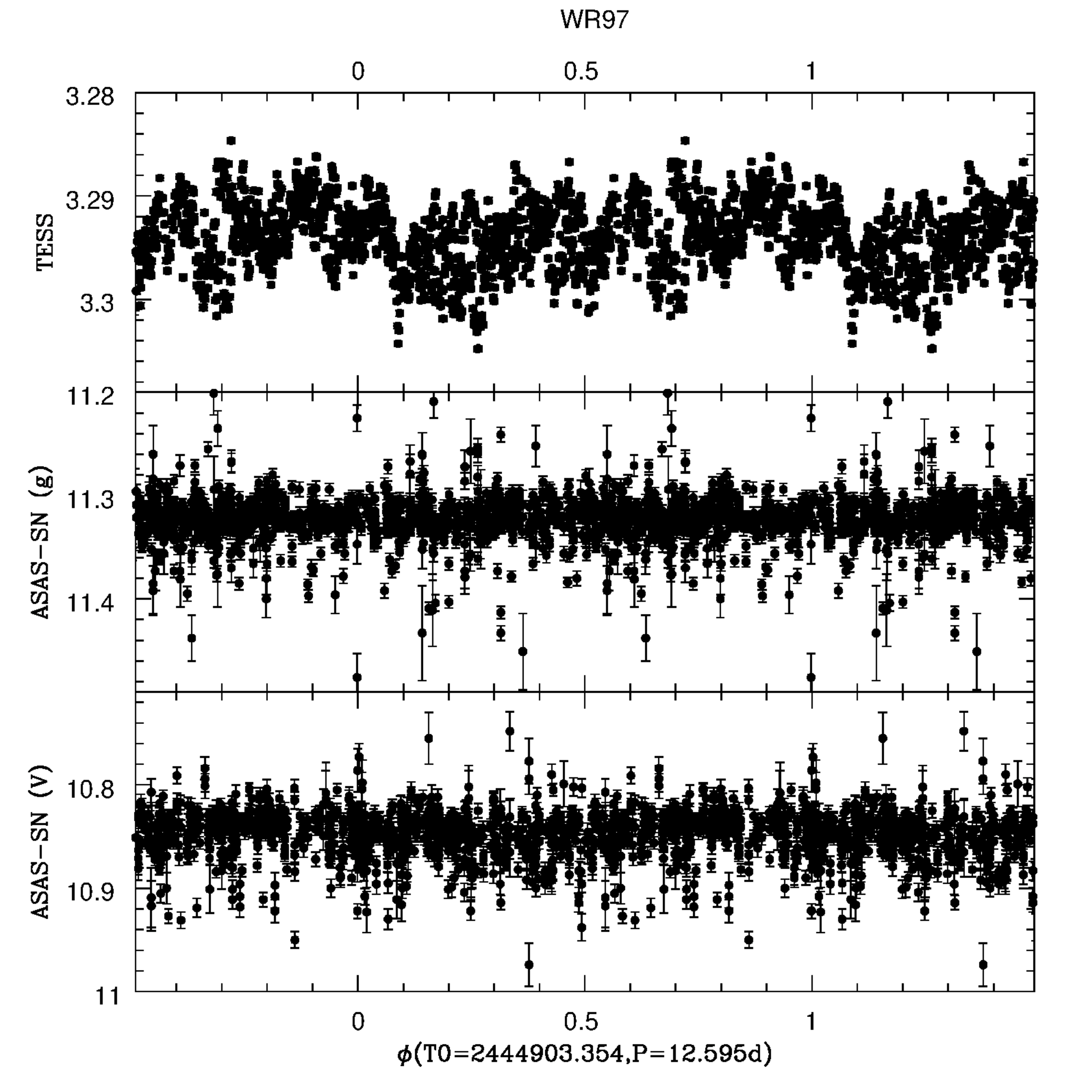}
  \end{center}
  \caption{Optical lightcurves of the \xmm\ targets, folded with ephemerides from Table \ref{journal}.}
\label{phot}
\end{figure*}
\setcounter{figure}{0}
\begin{figure*}
  \begin{center}
\includegraphics[width=5.8cm]{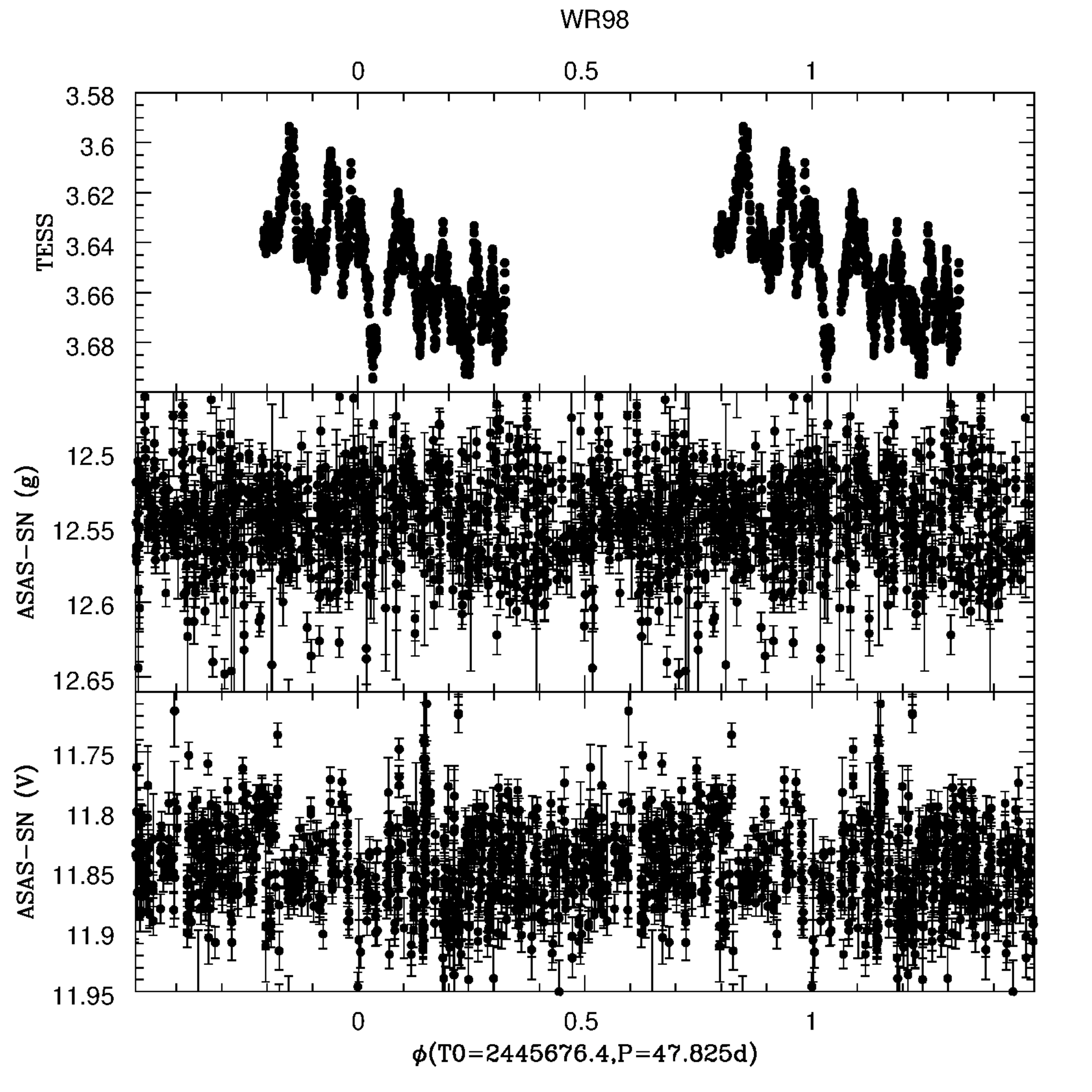}
\includegraphics[width=5.8cm]{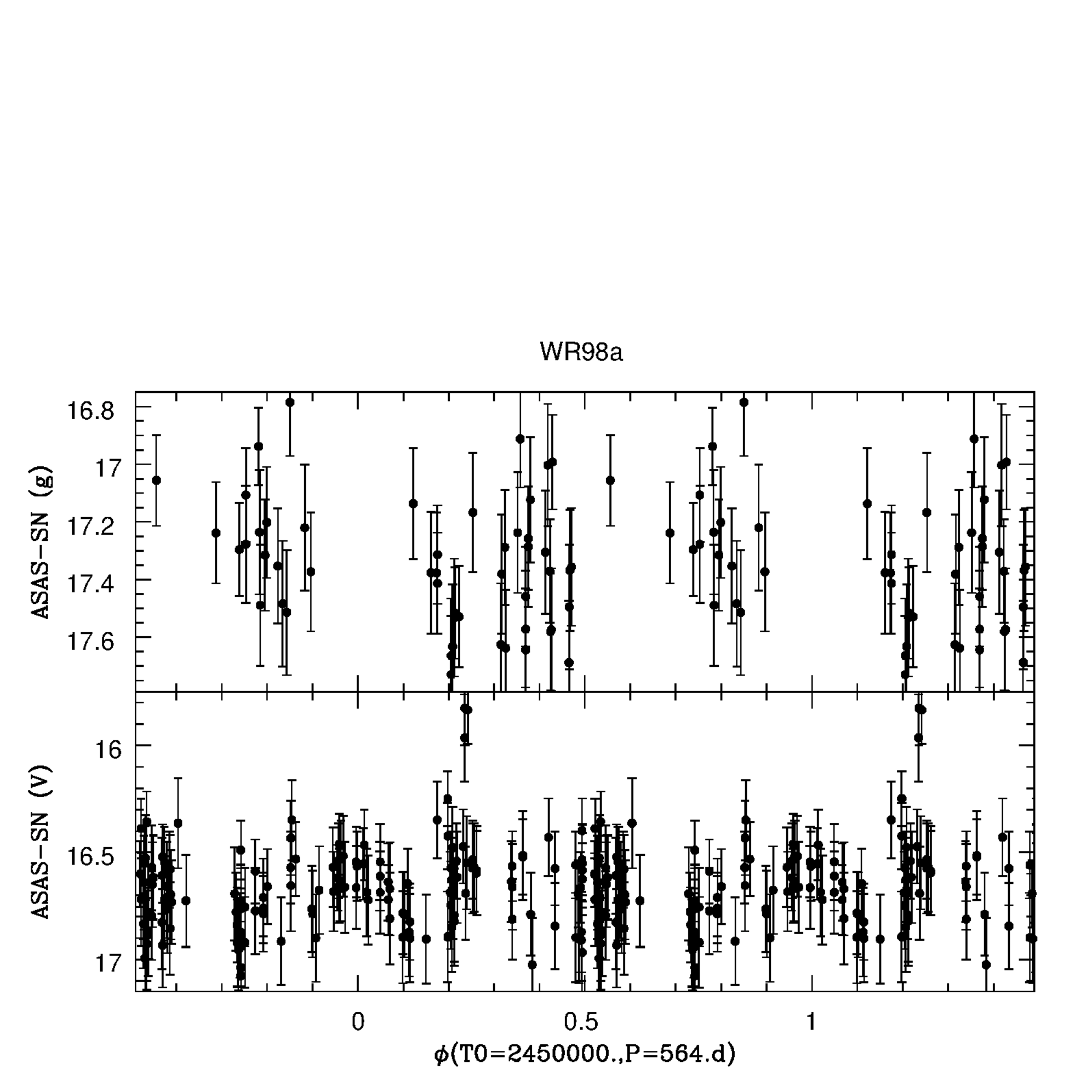}
\includegraphics[width=5.8cm]{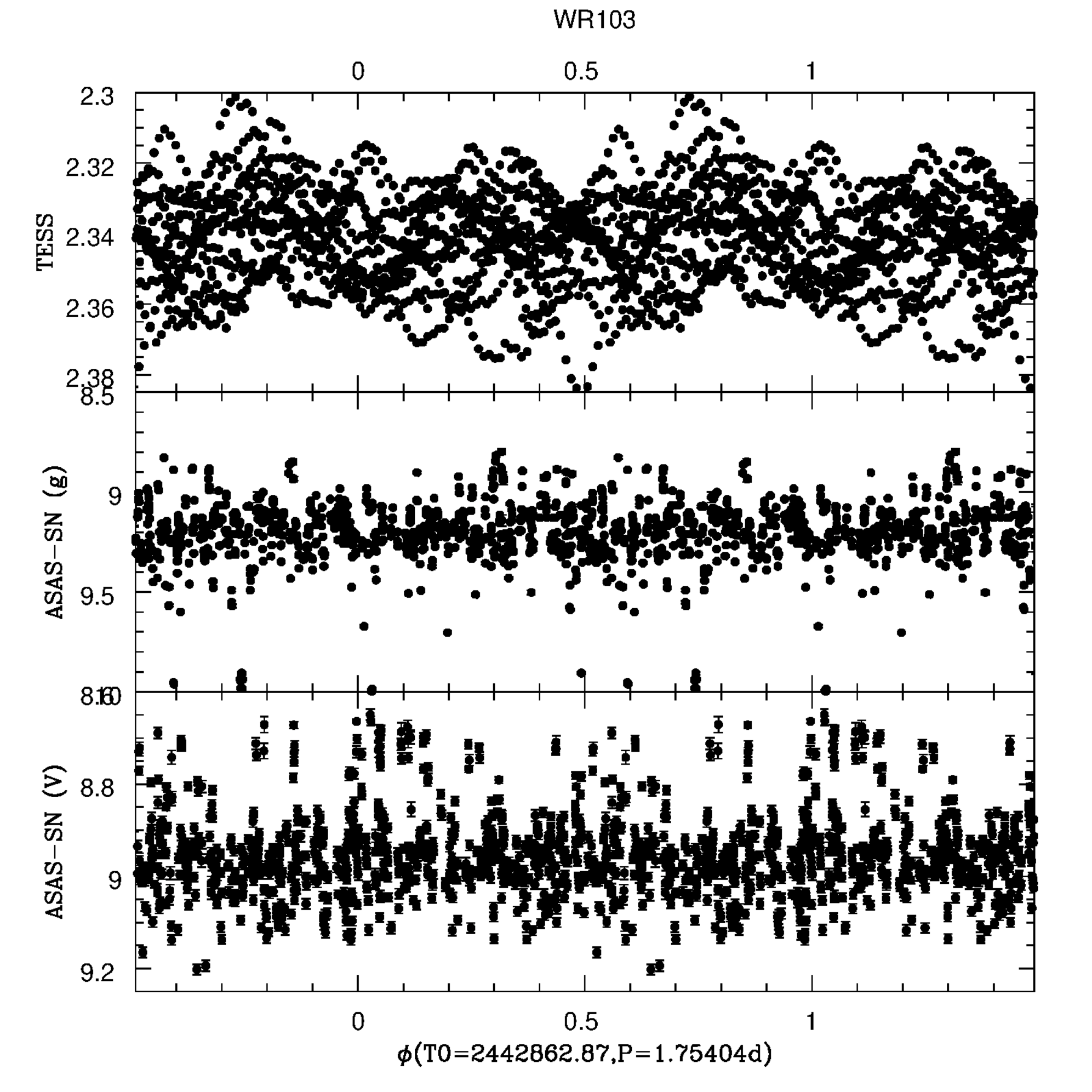}
\includegraphics[width=5.8cm]{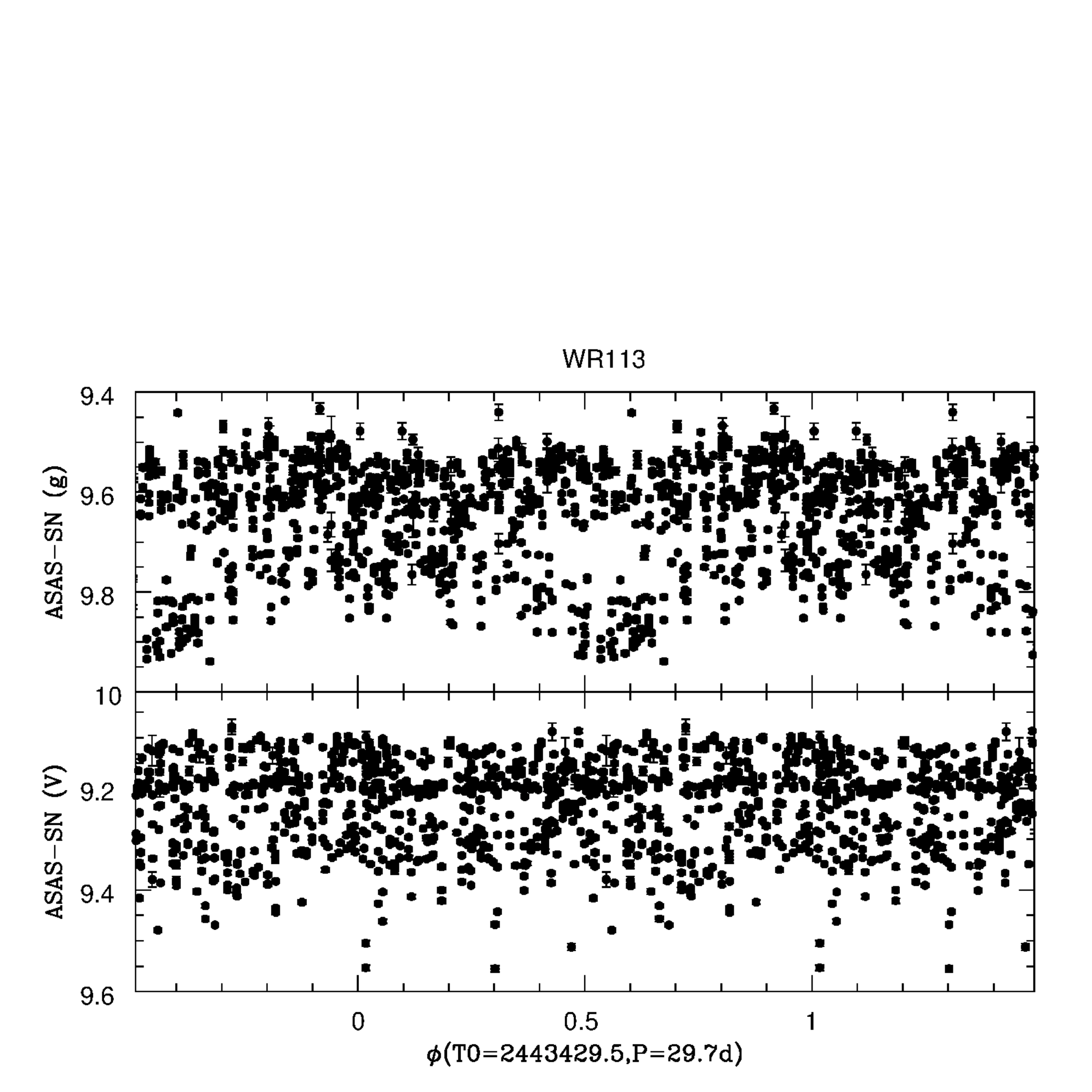}
\includegraphics[width=5.8cm]{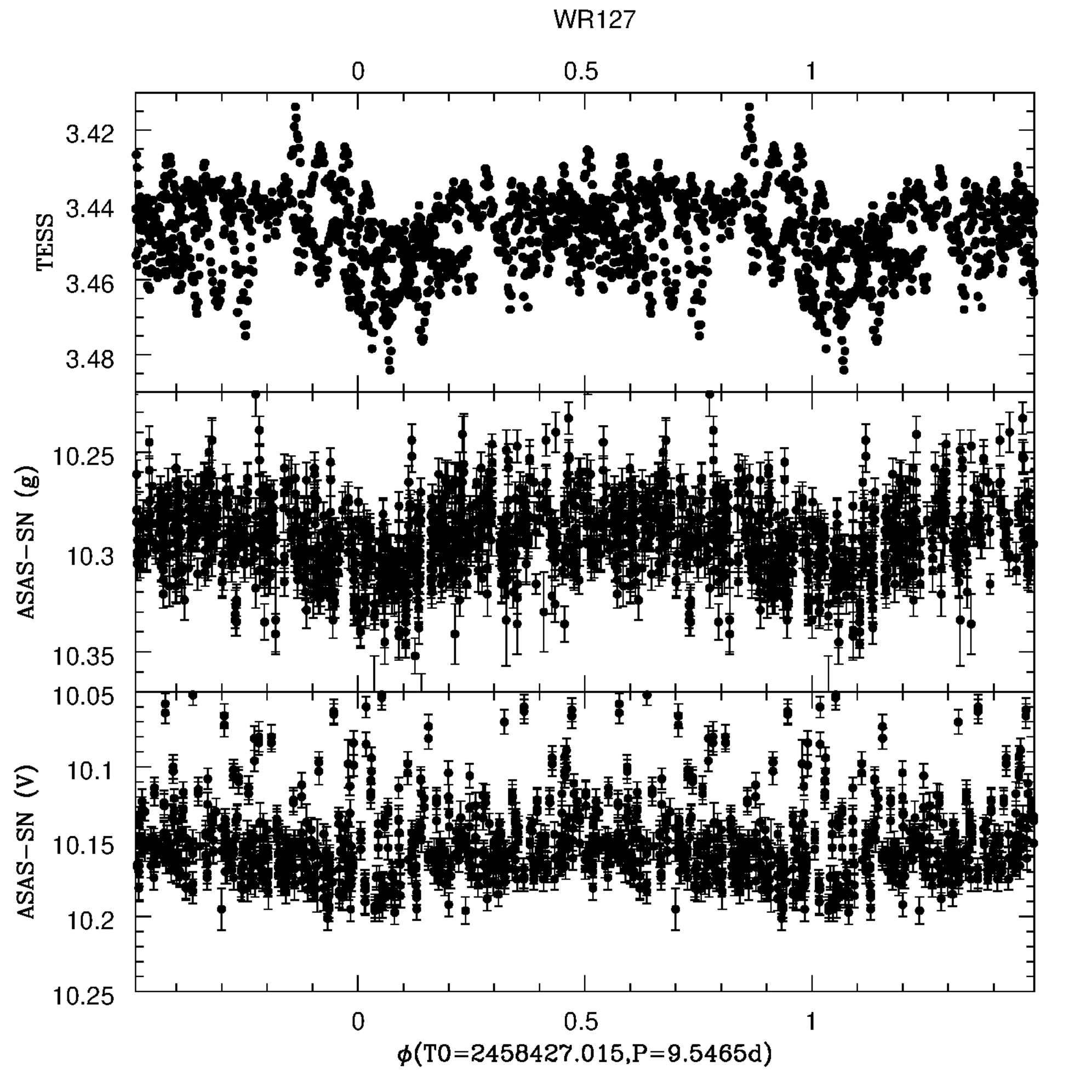}
\includegraphics[width=5.8cm]{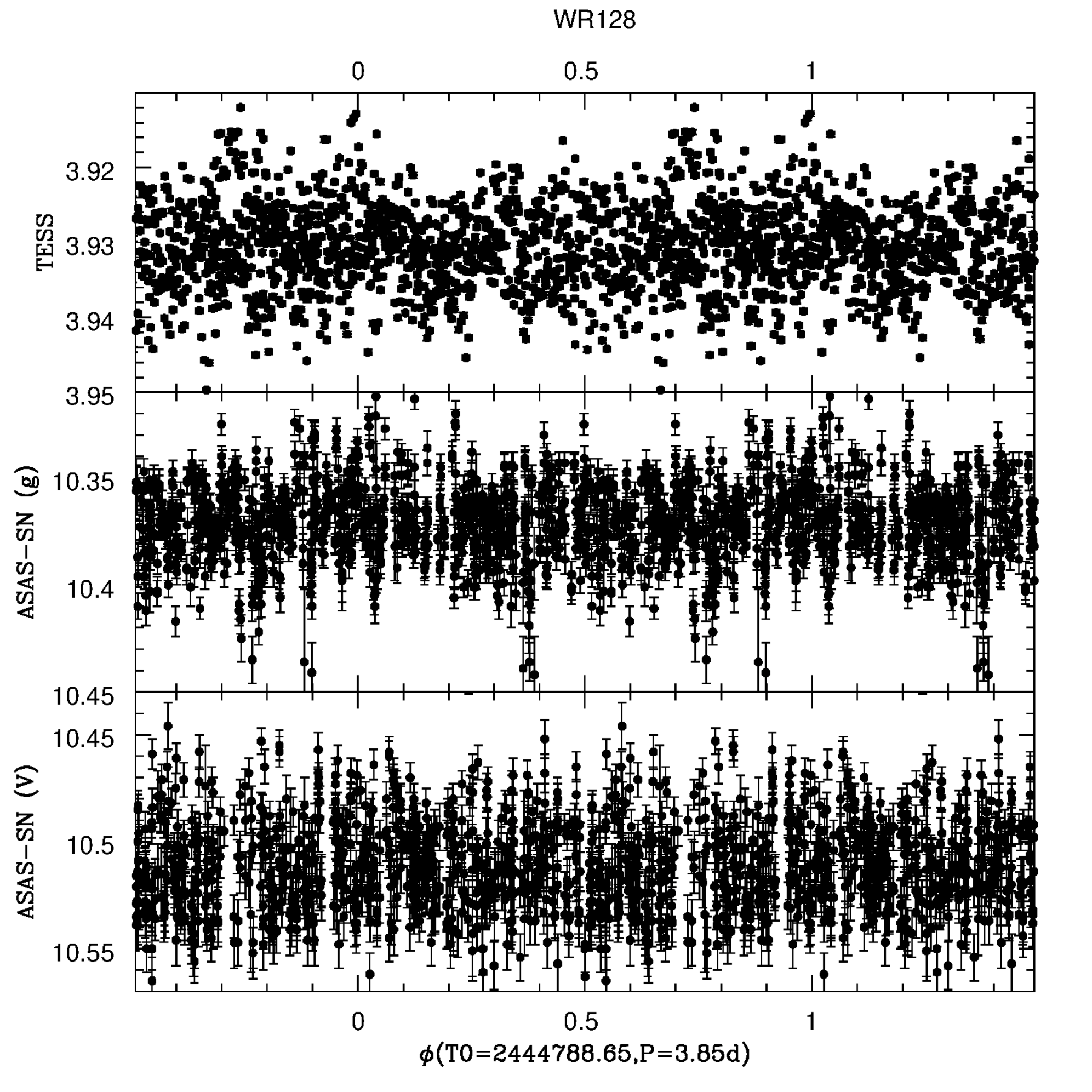}
\includegraphics[width=5.8cm]{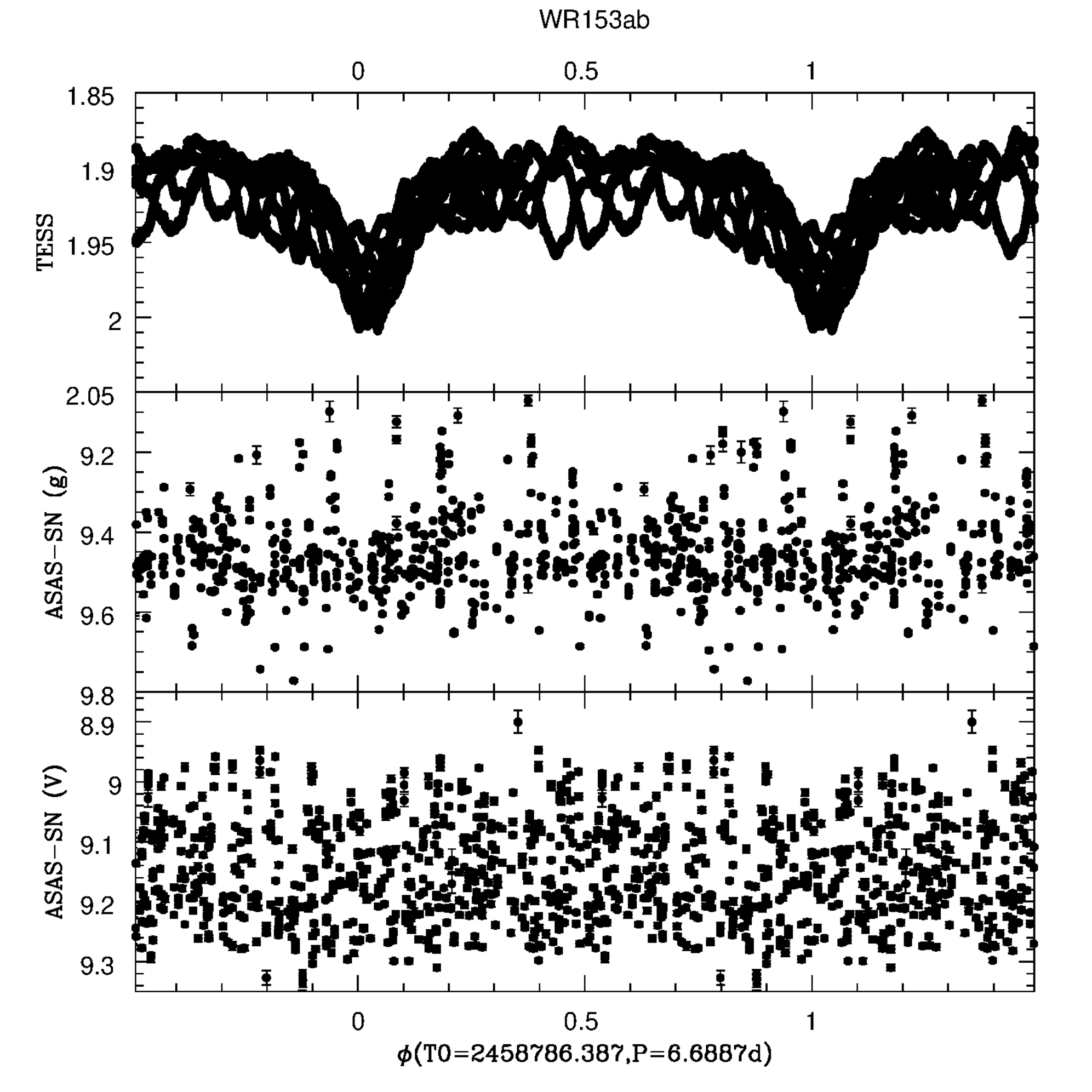}
\includegraphics[width=5.8cm]{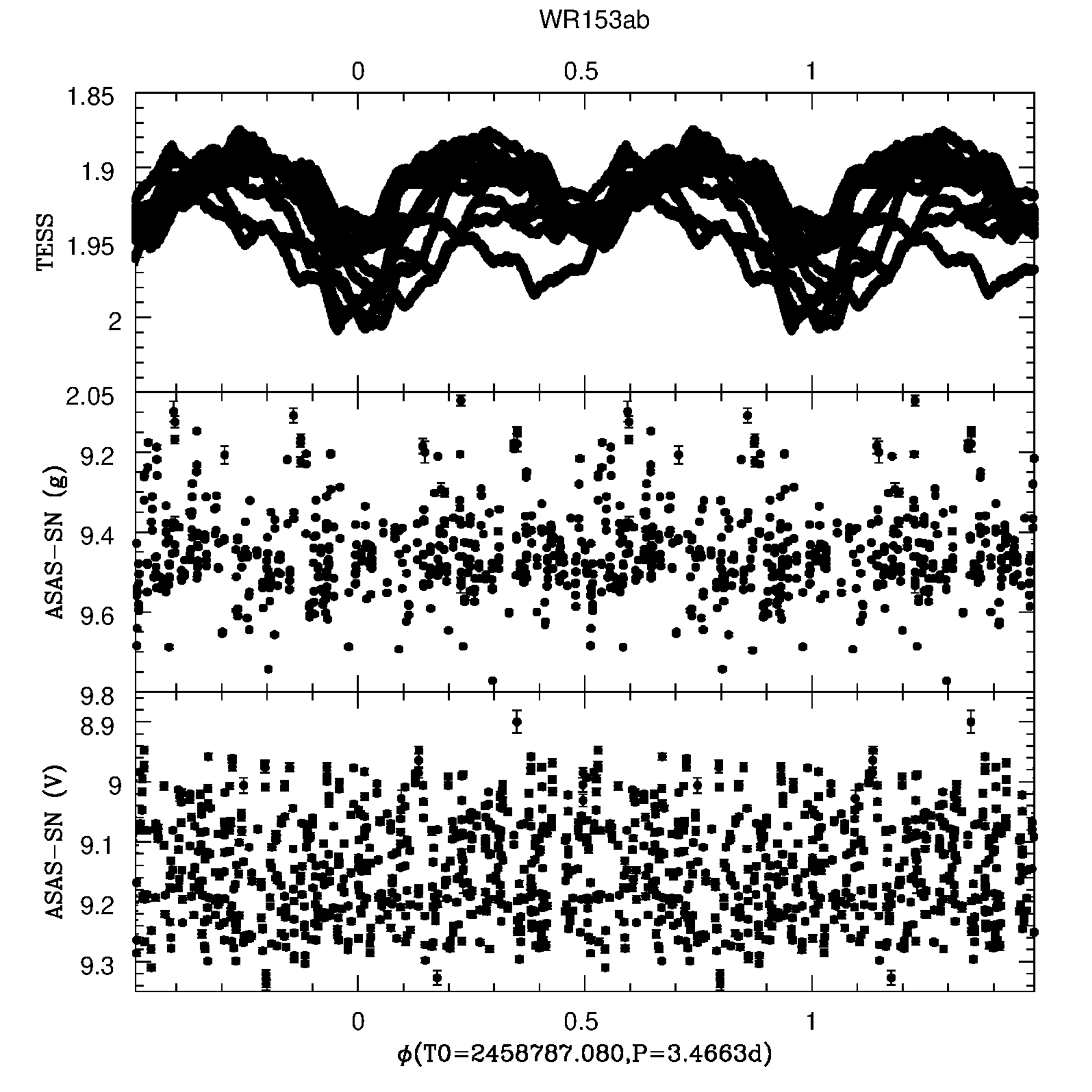}
  \end{center}
  \caption{Continued}
\end{figure*}

\begin{figure}
  \begin{center}
\includegraphics[width=7cm]{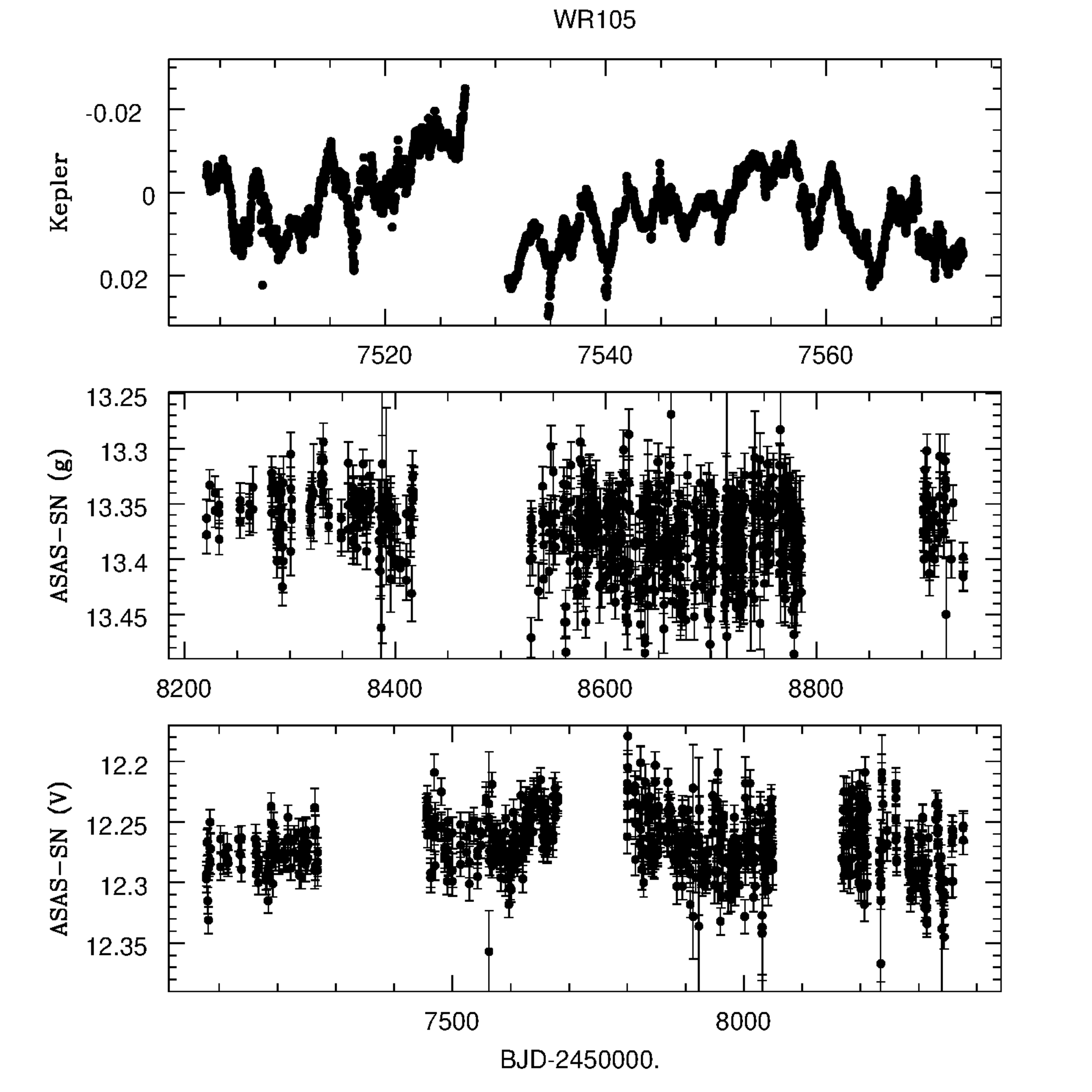}
  \end{center}
  \caption{Optical lightcurves of WR\,105, as a function of time }
\label{phot2}
\end{figure}

\bsp	% typesetting comment
\label{lastpage}
\end{document}